\documentclass[12pt]{iopart}
\pdfoutput=1

\usepackage{iopams}
\expandafter\let\csname equation*\endcsname\relax
\expandafter\let\csname endequation*\endcsname\relax
\usepackage{amsmath,amssymb,amsfonts}
\usepackage{graphicx,float,color}
\usepackage{bbold,bbm,cite}
\usepackage{tikz,drawInt}

\usepackage{etoolbox}
\makeatletter
\def\@mkboth#1#2{}
\newlength\appendixwidth
\preto\appendix{\addtocontents{toc}{\protect\patchl@section}}
\newcommand{\patchl@section}{%
  \settowidth{\appendixwidth}{\textbf{Appendix }}%
  \addtolength{\appendixwidth}{1.5em}%
  \patchcmd{\l@section}{1.5em}{\appendixwidth}{}{\ddt}%
}
\makeatother

\usepackage[colorlinks,bookmarks=false,citecolor=orange,linkcolor=blue,hyperfootnotes=true,urlcolor=blue]{hyperref}

\begin{document}

\title{Light cone dynamics in excitonic states of two-component Bose and Fermi gases}

\author{Neil J Robinson, Jean-S\'ebastien Caux}
\address{Institute for Theoretical Physics, University of Amsterdam, Postbus 4485, 1090 GL Amsterdam, The Netherlands}

\author{Robert M Konik}
\address{Condensed Matter Physics and Materials Science Division, Brookhaven National Laboratory, Upton, NY 11973-5000, USA}

\date{\today}

\begin{abstract}
We consider the non-equilibrium dynamics of two-component one dimensional quantum gases in the limit of extreme population imbalance where the minority species has but a single particle.   We consider the situation where the gas is prepared in a state with a single spatially localized exciton: the single particle of the minority species is spatially localized while the density of the majority species in the vicinity of the minority particle sees a depression.  Remarkably, we are able to consider cases where the gas contains on the order of $N=100$ particles, comparable to that studied in experiments on cold atomic gases.  We are able to do by exploiting the integrability of the gas together with the observation that the excitonic state can be constructed through a simple superposition of exact eigenstates of the gas.  The number of states in this superposition, rather than being exponentially large in the number of particles, scales linearly with $N$.  

We study the evolution of such spatially localized states in both strongly interacting Bose and Fermi gases.  The behavior of the light cones when the interaction strength and density of the gas is varied can be understood from exact results for the spin excitation spectrum in these systems.  We argue that the light cone in both cases exhibits scaling collapse.  However unique to the Bose gas, we show that the presence of gapped finite-momentum roton-like excitations provide the Bose gas dynamics with secondary light cones. 

\end{abstract}

\maketitle

\tableofcontents
\newpage

\section{Introduction}

Over the last decade, much work has focused on predicting and understanding the non-equilibrium dynamics of strongly correlated systems~\cite{polkovnikov2011colloquium,eisert2015quantum,dalessio2016from,cazalilla2016quantum,bernard2016conformal,essler2016quench,calabrese2016quantum,vidmar2016generalized,ilievski2016quasilocal,vasseur2016nonequilibrium,deluca2016equilibration,caux2016quench,langen2016prethermalization}. Integrable quantum systems have played a key role~\cite{essler2016quench,calabrese2016quantum,caux2016quench,ilievski2016quasilocal,vasseur2016nonequilibrium,deluca2016equilibration}: they can be exactly solved using well-known techniques~\cite{KorepinBook,GaudinBook} that, in some cases, allow the non-equilibrium dynamics to be computed analytically~\cite{calabrese2007quantum,calabrese2011quantum,calabrese2012quantum,schuricht2012dynamics,essler2012dynamical,collura2013equilibration,caux2013time,fagotti2014conservation,kormos2014analytic,essler2016quench,calabrese2016quantum,caux2016quench,ilievski2016quasilocal,vasseur2016nonequilibrium,deluca2016equilibration}. These studies have highlighted the important role of conservation laws in constraining the dynamics and relaxation of such systems, ultimately leading to an absence of thermalization~\cite{rigol2006hardcore,cazalilla2006effect,rigol2007relaxation,cramer2009exact,rigol2008thermalization,barthel2008dephasing,rigol2009breakdown,rossini2010long,fioretto2010quantum,fagotti2013reduced,mussardo2013infinitetime,pozsgay2013generalized,fagotti2013stationary,wouters2014quenching,pozsgay2014correlations,fagotti2014relaxation,denardis2014solution,goldstein2014failure,rigol2014quantum,sotiriadis2014validity,essler2015generalized,ilievski2015quasilocal,ilievski2015complete,calabrese2007quantum,calabrese2011quantum,caux2012constructing,calabrese2012quantum,schuricht2012dynamics,essler2012dynamical,collura2013equilibration,caux2013time,fagotti2014conservation,kormos2014analytic,essler2016quench,calabrese2016quantum,caux2016quench,ilievski2016quasilocal,vasseur2016nonequilibrium,deluca2016equilibration}. Instead, expectation values of local operators relax to time-independent values described by a generalized Gibbs ensemble~\cite{rigol2006hardcore,cazalilla2006effect,rigol2007relaxation,cramer2009exact,rigol2008thermalization,barthel2008dephasing,rigol2009breakdown,rossini2010long,fioretto2010quantum,fagotti2013reduced,mussardo2013infinitetime,pozsgay2013generalized,fagotti2013stationary,wouters2014quenching,pozsgay2014correlations,fagotti2014relaxation,denardis2014solution,goldstein2014failure,rigol2014quantum,sotiriadis2014validity,essler2015generalized,ilievski2015quasilocal,ilievski2015complete,calabrese2007quantum,calabrese2011quantum,caux2012constructing,calabrese2012quantum,schuricht2012dynamics,essler2012dynamical,collura2013equilibration,caux2013time,fagotti2014conservation,kormos2014analytic,essler2016quench,calabrese2016quantum,caux2016quench,ilievski2016quasilocal,vasseur2016nonequilibrium,deluca2016equilibration}. Such behavior can persist on intermediate time scales even when integrability is broken~\cite{langen2016prethermalization,moeckel2008interaction,rosch2008metastable,moeckel2009realtime,kollar2011generalized,vandenworm2013relaxation,marcuzzi2013prethermalization,essler2014quench,nessi2014quantum,brandino2015glimmers,bertini2015prerelaxation,bertini2015prethermalization,menegoz2015prethermalization,babadi2015farfrom,bertini2016thermalization}.  

In this work we are going to focus on the non-equilibrium dynamics of one-dimensional two-component quantum gases.  Such systems are also routinely studied in experiments on cold atomic gases~\cite{lewenstein2007ultracold,cazalilla2011onedimensional,stamperkurn2013spinor,guan2013fermi}.  Cold atomic gases give rise to unprecedented levels of control and tunability, both in terms of isolation of the system from the environment and in the systematic control of strength of interactions and tunneling. Furthermore, precise control over the constituents of the gas has enabled the study of the `extreme imbalance' limit~\cite{schirotzek2009observation,nascimbene2009collective,palzer2009quantum,kohstall2012metastability,koschorreck2012attractive,catani2012quantum,spethmann2012dynamics,scelle2013motional,yan2019bose}, which can be pictured as a single distinguishable impurity (the minority species) interacting with a bath (the majority species). This scenario has also attracted significant theoretical attention~\cite{klein2007dynamics,johnson2011impurity,goold2011orthoganlity,johnson2012breathing,mathy2012quantum,massel2013dynamics,kantian2014competing,burovski2014momentum,gamayun2014kinetic,gamayun2014quantum,lychkovskiy2014perpetual,massignan2014polarons,knap2014quantum,lychkovskiy2015perpetual,volosniev2015realtime,akram2016numerical,robinson2016motion,CastelnovoPRA16,gamayun2016time,gamayun2017quench,lychkovskiy2018necessary,lychkovskiy2018quantum,mistakidis2019dissipative}.  

In particular, we are focus on {\it integrable} two component gases.  Because of the two components, it is typically necessary to solve a nested Bethe ansatz.  Until recently, integrable systems solvable via the \textit{nested Bethe ansatz} have not received much attention, with studies instead focusing on spin-$1/2$ chains, spinless fermions, or the Lieb-Liniger model. While nested systems are technically more difficult to deal with, this is now changing with both purely numerical approaches, generalizations of the quench action~\cite{caux2016quench}, and the quantum transfer matrix approach~\cite{klumper1992free,klumper1993thermodynamics} tackling nested (i.e., multi-component) problems~\cite{robinson2016motion,mestyan2017exact,bertini2017quantum,piroli2019nested1,piroli2019nested2,modak2019nested}. Perhaps the simplest example of a model solved via the nested Bethe ansatz is the Yang-Gaudin model~\cite{yang1967some,gaudin1967systeme}, described by the Hamiltonian 
\begin{align}
  H = \int {\rm d}x\, \left[ \frac{\hbar^2}{2m} \sum_{\sigma = \uparrow,\downarrow} \partial_x \Psi^\dagger_\sigma(x) \partial_x \Psi_\sigma(x) + c \sum_{\sigma,\sigma' = \uparrow,\downarrow} \Psi^\dagger_{\sigma}(x) \Psi^\dagger_{\sigma'}(x) \Psi_{\sigma'}(x) \Psi_{\sigma}(x) \right].
\label{Eq:Ham}
\end{align}
This is a continuum theory of two species of particles that interact via an ultra-local delta function. Here $\sigma=\ \uparrow,\downarrow$ is a pseudo-spin index that labels the two species, $m$ is the particle mass, and $c$ is the strength of the interaction. The model is integrable for both fermionic and bosonic exchange statistics of the field $\Psi_\sigma(x)$ (or, indeed, any mixture of bosons or fermions)~\cite{yang1967some,gaudin1967systeme,sutherland1968further}. The Hamiltonian possesses an SU(2) symmetry, related to the presence of two species of equal mass. 

One of the main challenges to accessing the nonequilibrium dynamics in models with multiple species is a lack of knowledge of matrix elements of local operators in these systems. However, over the last few years there has been significant progress in obtaining efficient representations of matrix elements of local operators in the Yang-Gaudin model~\eqref{Eq:Ham}. The `extreme imbalance' limit was considered in Ref.~\cite{pozsgay2012on}, while advances in the study of form factors in GL(3)-invariant models~\cite{belliard2012algebraic,belliard2013form,pakuliak2014form,pakuliak2015zero,pakuliak2015form} have opened the door to obtaining matrix elements of local operators in more general cases~\cite{pakuliak2015form}. 

In this work, we focus on the extreme imbalance limit and will study quantum quenches in this limit.  In particular we want to treat a situation where the single particle of the minority component is initially localized and the subsequent time evolution of the gas is studied.   A challenge that we face in doing so is finding an efficient representation of this particular initial condition of the gas.  We can always find the initial condition numerically by expanding a gas with a localized particle (here taken to be $\downarrow$) in terms of the Bethe states (the eigenstates of the system as determined by the nested Bethe ansatz):
\begin{align}
&|{\rm N-particle~gas~with~one~\downarrow~particle~Gaussian~localized}\rangle \nonumber\\
& \hskip 0.1in = \frac{1}{\cal N} \int^{L/2}_{-L/2} {\rm d} x\, e^{-\alpha x^2}\psi^\dagger_\downarrow (x)|{\rm uniform~background~of~}N-1\,\uparrow\,{\rm particles}\rangle, \nonumber \\
& \hskip 0.1in =  \sum^K_{s=1} c_s |{\rm Bethe~states}\rangle.
\label{Eq:Init}
\end{align}
This requires that we employ the recently derived matrix elements discussed above in large numbers.  This number, $K$, grows exponentially in the number of particles, N, in the system.  Beyond $N\sim 10$, the number of needed matrix elements ($K^2$) to then track the subsequent temporal dynamics becomes numerically prohibitive.  While gases with $N\sim 10$ exhibit extremely rich behavior (such as the stuttering motion discussed in \cite{robinson2016motion}), there is a desire to study situation when the number of particles in the gas, if not approaching the thermodynamic limit, at least approximates what is found in experiments.

Remarkably we have found a simple representation of a state in the two component gas where the minority species particles is localized {\it and} where the number $K$ of states appearing in the representation of the initial state scales linearly with the number $N$ of particles in the gas.  That such a construction was even possible for 1d quantum gases was demonstrated in Refs. \cite{sato2012exact,kaminishi2011exact} for the bosonic Lieb-Liniger model (a one component gas) where `dark soliton' states were constructed.  In this work we generalize this construction to a two-component gas for both the cases of bosonic and fermionic particles.  We show that it is possible to study gases with relatively large numbers of particles  ($N = 100-1000$) compared to previous studies.

Our primary focus here will be on studying the light cone structure that results upon evolution of a quantum state with a localized impurity, such as that given in Eq.~\eqref{Eq:Init}.  We argue that both for fermions and bosons, the light cone exhibits scaling collapse as a function of the interaction parameter $\gamma$.  This scaling collapse arises because the excitation that defines the leading edge of the light cone, a spin excitation, exhibits scaling as a function of $\gamma$.  While both the fermionic and bosonic gases possess scaling, the bosonic gas possesses a secondary light cone.  We trace this difference to the spin excitation dispersion in the two cases.  In the fermionic case the dispersion has a single minima as a function of momentum, while in the bosonic case, there are multiple local minima (i.e. there is a `roton' in the bosonic dispersion \cite{li2003exact,robinson2017excitations}).  We argue that the secondary light cone arises because of the presence of these additional minima. It is worthwhile emphasizing that his secondary light cone is distinct from the primary light cone, i.e. we are not simply seeing a light cone with a broad edge as, say, observed in \cite{bertini2016thermalization}.

Our paper is organized as follows. In Sec.~\ref{Sec:Model} we first detail the solution of the Yang-Gaudin model~\eqref{Eq:Ham} for bosons and fermions via the Bethe ansatz, and discuss some of the known properties of the excitation spectra.  This is necessary, as indicated above, to understand the origin of the light cone behavior of the gases.  Following this, we introduce our initial states in Sec.~\ref{Sec:InitialState} and discuss how we perform the time-evolution of observables in Sec.~\ref{Sec:TimeEvo}. We then present results for the non-equilibrium dynamics emerging from the initial state in Sec.~\ref{Sec:NonEq}. We then discuss the aforementioned differences between the Fermi and Bose gases, originating in the spectrum of spin excitations. We conclude in Sec.~\ref{Sec:Conc} and present some complementary data in the Appendices. 

\section{Bethe ansatz solution of the Yang-Gaudin Model}
\label{Sec:Model}
In this section we discuss the Bethe ansatz solution of the Yang-Gaudin model~\eqref{Eq:Ham}. We first consider the case where the constituent particles are fermions, Sec.~\ref{Sec:Model:Fermions}, after which we consider the bosonic case in Sec.~\ref{Sec:Model:Bosons}. Independent of the exchange statistics of the particles, we consider repulsive interactions $c > 0$ and we impose periodic boundary conditions, placing a system of $N$ particles on a ring of circumference $L$. It will be useful to introduce the dimensionless interaction parameter
\begin{align}
  \gamma = \frac{2mc}{\hbar^2 \varrho}, \label{defGamma}
\end{align}
where $\varrho = N/L$ is the average particle density. Herein we will work in units where $\hbar = 2m = 1$.  

\subsection{The Yang-Gaudin Fermi gas}
\label{Sec:Model:Fermions}
Consider the Hamiltonian~\eqref{Eq:Ham} with spin-1/2 fermion fields
\begin{align}
  \big\{ \Psi_\sigma(x), \Psi^\dagger_{\sigma'}(y) \big\} = \delta_{\sigma,\sigma'} \delta(x-y), 
\end{align}
with $\sigma=\,\uparrow,\,\downarrow$ labeling an internal (pseudo)spin-1/2 degree of freedom. The Hamiltonian is integrable and the spectrum of eigenvectors and eigenvalues may be obtained via the nested Bethe ansatz~\cite{yang1967some,gaudin1967systeme,sutherland1968further,GaudinBook}. 

\subsubsection{Logarithmic Bethe ansatz equations.}
An eigenstate containing $N$ particles, of which $M$ are spin down, is denoted by $|\{p\};\{\lambda\}\rangle$. It is characterized by two sets of quantum numbers: a set of $N$ quasi-momenta $\{p\} \equiv \{p_1,\ldots,p_N\}$ and a set of $M$ spin rapidities $\{\lambda\} \equiv \{ \lambda_1 , \ldots, \lambda_M \}$. These satisfy a set of coupled non-linear algebraic equations~\cite{yang1967some,gaudin1967systeme,sutherland1968further,GaudinBook}, known as the logarithmic Bethe ansatz equations
\begin{eqnarray}\label{Eq:BAferms}
  2\pi \tilde I_j &= p_j L + \sum_{\alpha=1}^M \phi_2(p_j - \lambda_\alpha),\\
  2\pi \tilde J_\alpha &= \sum_{j=1}^N \phi_2(\lambda_\alpha-p_j) - \sum_{\beta=1}^M\phi_1(\lambda_\alpha-\lambda_\beta).
 \end{eqnarray}
 Here $\tilde I_j$, $\tilde J_\alpha$ are (half)integers satisfying 
\begin{eqnarray}
  \tilde I_j &\in& \left\{ \begin{array}{ll} \mathbb{Z}, \qquad & \mathrm{if~} M \in 2\mathbb{Z}, \\ \mathbb{Z}+\frac{1}{2}, \qquad & \mathrm{if~} M \in 2\mathbb{Z}+1, \end{array} \right. \\ 
  \tilde J_\alpha &\in& \left\{ \begin{array}{ll} \mathbb{Z}, \qquad & \mathrm{if~} (N+M) \in 2\mathbb{Z}+1, \\ \mathbb{Z}+\frac12, \qquad & \mathrm{if~} (N+M) \in 2\mathbb{Z}, \end{array} \right. \\
  \tilde I_i &\neq& \tilde I_j \quad {\rm if~} i\neq j, \qquad \tilde J_\alpha \neq \tilde J_\beta \quad {\rm if~} \alpha\neq\beta, \nonumber
\end{eqnarray}
and $\phi_n(u)$ is the scattering phase
\begin{align}
  \phi_n(u) = i \log \left(\frac{ \frac{ic}{n} + u}{\frac{ic}{n} - u} \right) \equiv 2\arctan\Big(\frac{un}{c}\Big).
  \label{Eq:scatphase}
\end{align}
The scattering phase~\eqref{Eq:scatphase} is bounded $|\phi_n(u)| \le \pi/2$ and hence so are the integers $|\tilde J_\alpha| \le (N+M)/2$. An analogous restriction on the (half)integers applies in the one-dimensional Hubbard model~\cite{HubbardBook}. 

Eigenstates $|\{p\};\{\lambda\}\rangle$ have momentum $P_{\{p\}}$ and energy $E_{\{p\}}$ that are expressed solely in terms of the quasi-momenta:
\begin{eqnarray}
  P_{\{p\}} &=& \sum_{j=1}^N p_j = \frac{2\pi}{L} \Bigg( \sum_{j=1}^N \tilde I_j - \sum_{\alpha=1}^M \tilde J_\alpha \Bigg), \label{Eq:momentum}\\
  E_{\{p\}} &=&\sum_{j=1}^N p_j^2\ . \label{Eq:energy}
\end{eqnarray}
The integrability of the model~\eqref{Eq:Ham} is realized in the constants of motion, which can also be expressed in terms of the quasi-momenta, ${\cal I}_n = \sum_{j=1}^N p_j^n$ with $n\in\mathbb{N}$. The absence of the spin rapidities in each of these expressions is why $\{\lambda_1,\ldots,\lambda_M\}$ are often called `auxiliary Bethe roots'. 

\subsubsection{Excitations of the Yang-Gaudin Fermi Gas.}
The Yang-Gaudin model~\eqref{Eq:Ham} has an SU(2) symmetry associated with the pseudo-spin index $\sigma$. The eigenstates of the model can be classified according to their pseudo-spin $S$, ranging from $S = 0$ to $S=N/2$. For the Fermi gas, the ground state is unpolarized~\cite{lieb1962theory}. Two types of excitations, related to the (half)integers $\tilde I_j$, $\tilde J_\alpha$ appearing within the logarithmic Bethe ansatz equations~\eqref{Eq:BAferms}, can be identified~\cite{yang1967some,gaudin1967systeme,sutherland1968further,GaudinBook}. The first type of excitation, which we will call a \textit{charge excitation}, is constructed by creating particle-hole excitations upon the Fermi sea of integers $\tilde I_j$.  We illustrate one such particle-hole excitation for $N=6$ particles below:  
\begin{center}
        \halfintegers
        \border
        \intrange{-5}{6}
        \fillI{-2}{3}
        \holeadd{1}{}
        \particleadd{5}{}
        \intLabel{ \left\{\tilde I\right\}}
        \render
\end{center}
The second type of excitation, which we call `spin' excitations, correspond to the presence of integers $\tilde J_\alpha$. For $M > 1$ the spin rapidities can organize into strings in the complex plane, corresponding to bound states of spin excitations. 

For models possessing an SU(2) symmetry, general considerations~\cite{halperin1969hydrodynamic,halperin1975dynamic} impose that spin excitations \textit{above the fully polarized state} are gapless and have quadratic dispersion at small momentum $p \ll \varrho$. 
\begin{align}
  E_s(p) \approx \frac{p^2}{2m^*}.
\end{align}
The effective mass $m^\ast$ for the spin excitations can become very large ($m^\ast \sim N$) at strong coupling $\gamma\to\infty$ due to the hard-core interaction~\cite{mcguire1965interacting}.

\subsubsection{Properties of the Yang-Gaudin Fermi gas.}
As one of the simplest examples of an integrable, interacting multi-component quantum gas, the Yang-Gaudin Fermi gas has been extensively studied, starting from its exact solution in 1967~\cite{yang1967some,gaudin1967systeme}. Equilibrium properties of the model are well understood, see the review~\cite{guan2013fermi}; the zero-temperature ground state was shown to be unpolarized by Lieb and Mattis~\cite{lieb1962theory}, with both the ground state and low-energy excitations being studied in detail by Schlottmann in the mid-1990s~\cite{schlottmann1993thermodynamics,schlottmann1994ground}. Recently, the finite-temperature thermodynamic properties were also consider, see Ref.~\cite{patu2016thermodynamics}. Some of the interest in this model comes from the fact that the strong coupling limit $c\to\infty$ describes the low-density limit of the Hubbard model~\cite{HubbardBook}. 

Experiments on cold atomic gases have realized the Yang-Gaudin Fermi gas. The phase diagram at finite polarization was studied, both experimentally and via the Bethe ansatz, in Ref.~\cite{liao2010spin}. This revealed that the spin-imbalanced Yang-Gaudin Fermi gas is a prime candidate for the observation of Fulde-Ferrell-Larkin-Ovchinnikov superconductivity~\cite{fulde1964superconductivity,larkin1965inhomogeneous}. The fully-polarized state has also been prepared experimentally, and was used to study deviations from the Yang-Gaudin model due to $p$-wave scattering~\cite{gunter2005pwave}. Cold atomic gas experiments have also studied the formation of bound states~\cite{moritz2005confinement} and the addition of an optical lattice~\cite{heinze2013intrinsic}.

\subsection{The Yang-Gaudin Bose gas}
\label{Sec:Model:Bosons}
Let us now consider the Yang-Gaudin model~\eqref{Eq:Ham} with two-component Bose fields
\begin{align}
  \big[ \Psi_\sigma(x), \Psi^\dagger_{\sigma'}(y) \big] = \delta_{\sigma,\sigma'} \delta(x-y),
\end{align}
where the index $\sigma=\uparrow,\downarrow$ is an internal label for the two species of bosons. In experiments on cold atomic gases, this may index two hyperfine states~\cite{palzer2009quantum,nicklas2015observation} or two different species of atom~\cite{catani2012quantum}. 

\subsubsection{Logarithmic Bethe ansatz equations.}
As with the Fermi gas, the Yang-Gaudin Bose gas is integrable and exactly solvable~\cite{yang1967some,sutherland1968further}. An eigenstate containing $N$ particles, of which $M$ are spin down, is denoted by $|\{k\};\{\nu\}\rangle$ and is characterized by a set of $N$ quasi-momenta $\{q\} \equiv \{q_1,\ldots,q_N\}$ and a set of $M$ spin rapidities $\{\nu\} \equiv \{\nu_1,\ldots, \nu_M\}$. These quantum numbers satisfy the following set of logarithmic Bethe ansatz equations~\cite{yang1967some,sutherland1968further} [cf. Eqs.~\eqref{Eq:BAferms}] 
\begin{eqnarray} \label{Eq:BAbosons}
2\pi I_i &= k_i L- \sum_{\alpha=1}^{M} \phi_2(k_i - \nu_\alpha) +\sum_{l=1}^N \phi_1(k_i - k_l) ,\\
2\pi J_\alpha &= \sum_{l=1}^N \phi_2 (\nu_\alpha - k_l) - \sum_{\beta=1}^{M} \phi_1(\nu_\alpha - \nu_\beta),
\end{eqnarray}
where $\phi_n(u)$ is defined in Eq.~\eqref{Eq:scatphase}. The (half) integers $I_i$, $J_\alpha$ satisfy
\begin{eqnarray}
I_i, J_\alpha &\in&  \left\{ \begin{array}{lcl} \mathbb{Z} & \quad & \mathrm{if~} (N+M) \in 2\mathbb{Z}+1, \label{Eq:QNs1} \\
\mathbb{Z} + \frac12 & \quad & \mathrm{if~} (N+M)\in 2\mathbb{Z}, \end{array} \right. \label{Eq:QNs2} \\
I_i \neq I_j && \text{if~} i\neq j, \qquad J_{\alpha} \neq J_{\beta} \quad \text{if~} \alpha\neq\beta. \label{Eq:BoseExcInt}
\end{eqnarray}
As in the fermionic case, bounding of the scattering phase~\eqref{Eq:scatphase} leads to bounding of the integer $|J_\alpha| \le (N+M)/2$.

Also in analogy with the Fermi Yang-Gaudin model, the energy $E_{\{k\}}$ and momentum $P_{\{k\}}$ of the $N$ particle eigenstates $|\{k\};\{\nu\}\rangle$ are given by
\begin{align}
P_{\{k\}} &= \sum_{j=1}^{N} k_j = \frac{2\pi}{L} \left( \sum_{j=1}^{N} I_j - \sum_{\alpha=1}^M J_\alpha \right), \\
E_{\{k\}} &= \sum_{j=1}^N k_j^2.
\end{align}
The ground state of the Bose gas is formed from a `Fermi sea' of integers $\{I\}$, symmetrically distributed about the origin. This follows directly from the exclusion principle for integers, Eq.~\eqref{Eq:BoseExcInt} and the monotonicity property of the scattering phase, Eq.~\eqref{Eq:scatphase}.

\subsubsection{Excitations of the Yang-Gaudin Bose Gas.}
We can once again define `charge' and `spin' excitations in the Bose gas, corresponding to particle-hole excitations in the Fermi sea of integers $\{I\}$ and the presence of integers $\{J\}$, respectively.  As already mentioned, the SU(2) symmetry of the model fixes the spin excitations above the ground state to have a gapless quadratic dispersion. The effective mass for the spin excitations in the Bose gas can become very large at strong coupling, as discussed in Refs.~\cite{fuchs2005spin,zvonarev2007spin}. 

\subsubsection{Properties of the Yang-Gaudin Bose Gas.}
The equilibrium properties of the Yang-Gaudin Bose gas are relatively well understood~\cite{stamperkurn2013spinor}. There is a well known theorem that the ground state of the two-component Bose has with spin-independent interaction (as is the case in the Yang-Gaudin model~\eqref{Eq:Ham}) is spin polarized~\cite{suto1993percolation,eisenberg2002polarization}; this is also seen from the Bethe ansatz solution~\cite{yang2003rigorous,li2003exact,robinson2017excitations}. The Bethe ansatz solution can also be used to study finite temperature properties, such as the behavior of dressed excitations~\cite{li2003exact,robinson2017excitations}, the temperature-dependent polarization~\cite{caux2009polarization}, and other thermodynamic properties~\cite{klauser2011equilibrium}. It can also be extended to the case of inhomogeneous systems, such as gases confined in a trap, using the local density approximation~\cite{patu2015thermodynamics}. 

Excitations above the zero-temperature spin-polarized ground state have been extensively studied~\cite{li2003exact,fuchs2005spin,robinson2017excitations}. At strong coupling, unusual `logarithmic diffusion' of spin excitations was observed in dynamical correlation functions due to their large effective mass~\cite{fuchs2005spin,zvonarev2007spin}. In the nonequilibrium context much less is understood, with progress being hampered by the lack of knowledge of matrix elements. For finite coupling strengths, the dynamics of an impurity immersed in a small bath ($N\sim10$ particles) was studied in Ref.~\cite{robinson2016motion}, revealing rather rich dynamics. 

\section{The initial state}
\label{Sec:InitialState}
Having introduced the Yang-Gaudin model~\eqref{Eq:Ham} and its exact solution via the Bethe ansatz, for both fermions and bosons, we will now discuss the initial states that we consider in our nonequilibrium dynamics. We focus on the `extreme imbalance' limit of the Yang-Gaudin model, constructing a state that contains a single localized flipped spin (i.e., $M=1$). Our choice of state is motivated by the construction of Sato \textit{et al.}~\cite{sato2012exact} for initially localized states in the Lieb-Liniger model. We first introduce this case and then discuss our generalization to the Yang-Gaudin model. 

\subsection{Initially localized states in the Lieb-Liniger model}
The Lieb-Liniger model describes a single species of delta-function interacting bosons, and it is solvable via the Bethe ansatz~\cite{lieb1963exact,lieb1963exactII,KorepinBook}. $N$ particle eigenstates on the ring of size $L$, denoted $|\{q\}\rangle$, have energy $E_{\{q\}} = \sum_j q_j^2$ and momentum $P_{\{q\}} = \sum_j q_j$. The set of quasi-momenta $\{q\} \equiv \{q_1,\ldots,q_N\}$ satisfy the logarithmic Bethe ansatz equations
\begin{align}
  q_j L = 2\pi {\cal I}_j - 2 \sum_{\ell \neq j} \arctan\left( \frac{q_j - q_\ell}{c} \right), \label{Eq:LLBAE}
\end{align}
where $c$ is the interaction strength of the delta function, and ${\cal I}_j$ are integers (half-odd integers) for $N$ odd (even). For further information about the exact solution, see, e.g., Ref.~\cite{KorepinBook}.  

In Ref.~\cite{sato2012exact}, Sato \textit{et al.} proposed a remarkably simple state $|X\rangle$ that contains a density dip at position $X + L/2$ on the ring:
\begin{align}
  |X \rangle = \frac{1}{\sqrt N} \sum_{p=0}^{N-1} \exp\left(-2\pi i \frac{pq}{N}\right) |p\rangle, \label{Eq:satostate}
\end{align}
where the states $|p\rangle$ are described by the set of quasi-momenta that are solutions of the Bethe equations~\eqref{Eq:LLBAE} with the configuration of integers
\begin{align}
  {\cal I}_j = \left\{ \begin{array}{lll}
                         -\frac{N+1}{2} + j & \quad & \text{for~} 1\le j \le N-p, \\
                         -\frac{N+1}{2} + j + 1 & \quad & \text{for~} N-p+1 \le j \le N.
                       \end{array}\right.
\end{align}
The integer $0 \le q \le N-1$ sets $X = q L/N$ and hence the position at which the density notch is initially localized. We note that states of the form $|X\rangle$ are \textit{not eigenstates of the Hamiltonian}, and hence time-evolution is non-trivial. 
 
In Refs.~\cite{sato2012exact,sato2012quantum,kaminishi2014exact,kaminishi2015recurrence} the states~\eqref{Eq:satostate} were used to study the dynamics of density notch (sometimes called a `dark soliton'). This included studies of how the notch dissolves in time, and the finite-size recurrences that occur during unitary time-evolution. The simple structure of the states~\eqref{Eq:satostate} means that the dynamics can be computed in a \textit{numerically exact manner} for large number of particles (e.g., $N\sim 1000$) with modest computational effort. 

\subsection{Initially localized spin excitations in the Yang-Gaudin model}
Motivated by the construction discussed above, we consider states formed from a \textit{linear superposition of eigenstates with a fixed set of quasi-momenta integers $\{I\}$ and a single spin integer $J$ varied between its bounds}: 
\begin{align}
  \big| \{I\} \big\rangle = \frac{1}{\sqrt{\cal N}}
  \sum_{|J| < \frac{N}{2}} \big| \{p\}_J; \lambda_J \rangle. \label{Eq:InitialState}
\end{align}
Here ${\cal N}$ is a normalization constant, and we introduce the notation $\{p\}_J$, $\lambda_J$ to denote the set of momenta and the spin rapidity obtained by solving the logarithmic Bethe ansatz equations (Eqs.~\eqref{Eq:BAferms} for fermions, Eqs.~\eqref{Eq:BAbosons} for bosons) with the set of integers $\{I\}$ and the single spin integer $J$. 

For the purpose of this work, we focus in particular on the Fermi sea configuration of the integers $\{I\}$. In this case, the initial state~\eqref{Eq:InitialState} can be pictured as a linear superposition of states containing a single spin excitations with a given momentum $P_{\{p\}_J}$~\cite{fuchs2005spin,zvonarev2007spin}. Note that in the Yang-Gaudin Fermi gas, such a state is far from the absolute ground state, which resides in the unpolarized $S=0$ sector. On the other hand, in the Bose gas, the constructed states with be close in energy to the fully polarized ground state. 

To be concrete, we restrict our attention to states containing $N$ even particles. For fermions, $I_j \in \mathbb{Z}+\frac12$ take half odd integer values, leading to a unique Fermi sea configuration
\begin{align}
  \{I\}^{(f)} = \left\{
  - \frac{N-1}2, -\frac{N-1}{2}+1, \ldots, \frac{N-1}{2}
  \right\}. \label{Eq:If}
\end{align}
We illustrate this configuration for $N=6$ particles below: 
\begin{center}
        \halfintegers
        \border
        \intrange{-4}{5}
        \fillI{-2}{3}
        \intLabel{ \left\{I\right\}^{(f)}}
        \render
\end{center}

On the other hand, for bosons the situation is slightly more complicated with $I_j \in \mathbb{Z}$ taking integer values. Thus there are two configurations for the Fermi sea
\begin{align}
  \{I\}^{(b)}_1 &= \left\{ - \frac{N}2, -\frac{N}2 + 1, \ldots, \frac{N}{2} - 1\right\}, \label{Eq:Ib1}\\
  \{I\}^{(b)}_2 &= \left\{ - \frac{N}{2}+1, - \frac{N}{2}  + 2, \ldots, \frac{N}{2} \right\}. \label{Eq:Ib2} 
\end{align}
We illustrate these two configuration for $N=6$ particles below: 
\begin{center}
        \integers
        \border
        \intrange{-5}{5}
        \fillI{-3}{2}
        \intLabel{ \left\{I\right\}^{(b)}_1}
        \render
\end{center}
\begin{center}
        \integers
        \border
        \intrange{-5}{5}
        \fillI{-2}{3}
        \intLabel{ \left\{I\right\}^{(b)}_2}
        \render
\end{center}
For the initial state~\eqref{Eq:InitialState} in the bosonic Yang-Gaudin gas, we construct the superposition with both of the configuration of integers $\{I\}^{(b)}_1$, $\{I\}^{(b)}_2$. 
 
We consider two further choices for the integers $\{I\}$, highlighting differences and similarities to the above, in \ref{App:diluted} and \ref{App:shifted}.

\subsubsection{Particle densities in the initial state.}
\label{Sec:Densities}
In proposing the initial state~\eqref{Eq:InitialState}, we were motivated by the construction of a density notch in the Lieb-Liniger model~\cite{sato2012exact,sato2012quantum,kaminishi2014exact,kaminishi2015recurrence}. Here we show that the state~\eqref{Eq:InitialState} with the Fermi sea of integers $\{I\}$ is a natural generalization of~\eqref{Eq:satostate}: there is a notch in the density of the majority $\sigma =\, \uparrow$ species, which is accompanied by a localized particle density in the minority $\sigma=\,\downarrow$ species. This is schematically reminiscent of an ``exciton'' particle-hole excitation. To see this, we compute the density of particles of species $\sigma=\,\uparrow,\downarrow$ within the state $|\{I\}\rangle$:
\begin{align}
\rho_\sigma(x) = \langle \{I\} | \Psi^\dagger_\sigma(x) \Psi_\sigma(x) | \{I\}\rangle. 
\end{align}

\begin{figure}[t]
\begin{center}
\begin{tabular}{cc}
\includegraphics[width=0.5\textwidth]{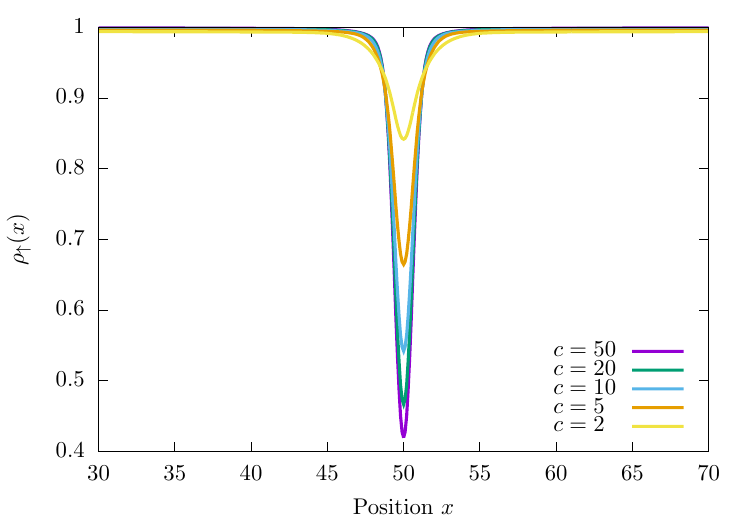}
\includegraphics[width=0.5\textwidth]{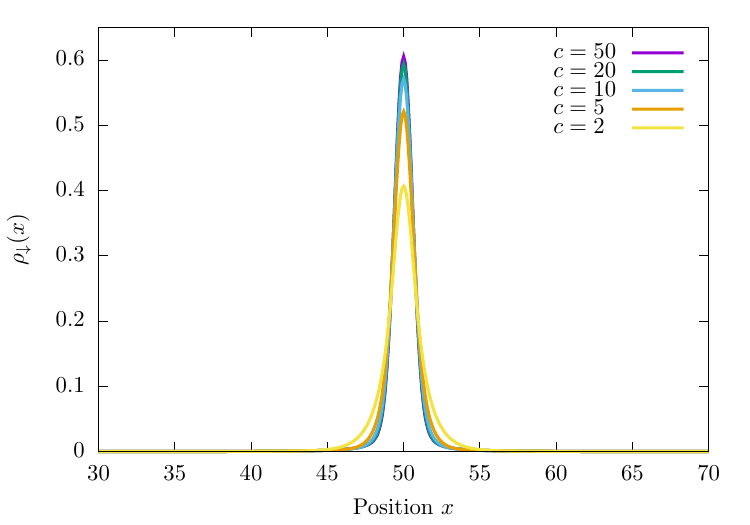}
\end{tabular}
\end{center}
\caption{The density $\rho_\sigma(x) = \langle \{I\}^{(f)}| \Psi^\dagger_\sigma(x) \Psi_\sigma(x)| \{I\}^{(f)} \rangle$ for (left panel) the majority $\sigma = \, \uparrow$; (right panel) the minority $\sigma =\, \downarrow$ fermions in the initial state~\eqref{Eq:InitialState} with the Fermi sea of integers $\{I\}^{(f)}$, see Eq.~\eqref{Eq:If}. The state is constructed for $N=100$ particles at unit filling $\varrho = 1$ for a range of interaction strengths, $c$.}
\label{Fig:InitialDensityFermions}
\end{figure}

We first consider the Yang-Gaudin Fermi gas in Fig.~\ref{Fig:InitialDensityFermions}, for both (a) the majority species $\sigma=\uparrow$ and (b) the minority species $\sigma=\downarrow$. In direct analogy to the Lieb-Liniger state~\eqref{Eq:satostate} of Ref.~\cite{sato2012exact}, we see that there is a clear density notch localized at $X=L/2$ in the majority species, Fig.~\ref{Fig:InitialDensityFermions}(a). The half-width at half-maximum of this notch decreases with increasing interaction strength (with a commensurate increase in depth), as also seen for the Lieb-Liniger state~\cite{sato2012exact}. Accompanying the density notch, the minority species is localized in a tight wave packet, of similar width to the notch, as shown in Fig.~\ref{Fig:InitialDensityFermions}(b). This localized excitation also narrows with increasing interaction strength. 

\begin{figure}[t]
\begin{center}
\includegraphics[width=0.6\textwidth]{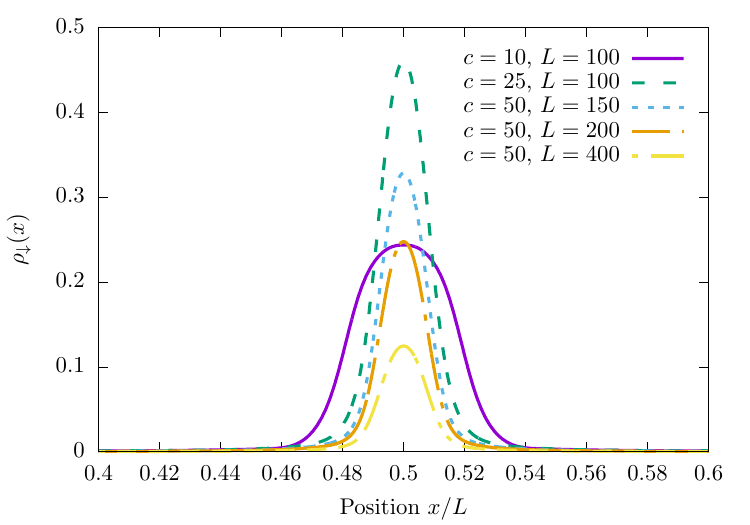}
\end{center}
\caption{The density profile $\rho_\downarrow(x)$ for the minority species of boson in the initial state~\eqref{Eq:InitialState} with the Fermi sea of integers $\{I\}^{(b)}$ [cf. Eqs.~\eqref{Eq:Ib1} and~\eqref{Eq:Ib2}]. The state is constructed with $N=100$ bosons on the ring of size $L$ for a number of interaction strengths.}
\label{Fig:InitialDensityBosons}
\end{figure}

On the other hand, we consider the Yang-Gaudin Bose gas. Here we are limited to studying only the minority species, $\sigma = \,\downarrow$, as an efficient representation for matrix elements of the majority density are currently not known to us (cf. Refs.~\cite{pozsgay2012on,pakuliak2015form}). Inferences about the majority species will be drawn in analogy with the Fermi gas, which we expect to be valid at strong coupling. In Fig.~\ref{Fig:InitialDensityBosons}, we see that there is a single well-localized wave packet of the minority particle, in direct analogy with the Fermi gas. We infer, in analogy with the Fermi gas, that there is an accompanying density notch in the background gas. Here the presented data is for $N=100$ boson on the ring of size $L=100,\, 150,\, 200,\, 400$ for a number of interaction strengths. We do note that there are some differences between the Bose and Fermi gases -- the Bose gas appears to be less localized than the Fermi gas for similar strength of the interaction. 

\section{Computing the time-evolution of observables}
\label{Sec:TimeEvo}

Before presenting our results for the nonequilibrium dynamics when starting from the initial state~\eqref{Eq:InitialState}, we briefly discuss how we compute the time-evolution of observables. Consider an arbitrary state $|\phi\rangle$. In the basis of eigenstates of the Hamiltonian~\eqref{Eq:Ham}, it may be written as 
\begin{align}
  |\phi\rangle = \sum_{\{p\},\{\lambda\}} c^\phi_{\{p\},\{\lambda\}} | \{p\};\{\lambda\}\rangle.
\end{align}
Here $\sum_{\{p\},\{\lambda\}}$ is shorthand for a sum over all states with quasi-momenta and spin rapidities satisfying the logarithmic Bethe ansatz equations. $c^\phi_{\{p\};\{\lambda\}} = \langle \{p\};\{\lambda\}| \phi \rangle$ are superposition coefficients (overlaps) that are normalized such that the wave-function has unit norm, $\langle \phi | \phi \rangle = 1$. In the basis of eigenstates, time-evolution is trivial: each eigenstate is acted upon by the time-evolution operator and picks up the appropriate phase factor 
\begin{align}
  e^{-iHt} |\{p\};\{\lambda\}\rangle = e^{-iE_{\{p\}}t}|\{p\};\{\lambda\}\rangle.
\end{align}
Formally, this gives us the time-evolved initial state
\begin{align}
  |\phi(t) \rangle \equiv e^{-iHt}|\phi\rangle = \sum_{\{p\},\{\lambda\}} c^\phi_{\{p\},\{\lambda\}} e^{-iE_{\{p\}}t} |\{p\};\{\lambda\}\rangle.
\end{align}

The computation of time-dependent expectation values can now be performed in a straight forward manner, provided we know matrix elements of the operator of interest. Consider the local operator $\hat O(x)$; the time-evolution of its expectation value is related to matrix elements of $\hat O(0)$ through  
\begin{align}
  \langle \phi (t) | \hat O(x) |\phi (t) \rangle =& \sum_{\{p\},\{\lambda\}} \sum_{\{k\},\{\mu\}} \Big(c^\phi_{\{p\},\{\lambda\}}\Big)^* c^\phi_{\{k\},\{\mu\}}  e^{it( E_{\{p\}} -E_{\{k\}})} e^{ix(P_{\{p\}} -P_{\{k\}})} \nonumber \\
  & \qquad\qquad \times \langle \{p\};\{\lambda\}|\hat O(0) | \{k\}; \{\mu\} \rangle\ .
\label{Eq:ExpectationValue}
\end{align}
Here we have used that the eigenstates $|\{k\};\{\mu\}\rangle$ are simultaneous eigenstates of the translation operator with eigenvalues $\exp(i P_{\{k\}} x)$. 

For an arbitrary state $|\phi\rangle$ undergoing time-evolution, the above expression~\eqref{Eq:ExpectationValue} is a double infinite sum---the Hilbert space of a continuum theory being infinite---and it is necessary to truncate these sums in some systematic manner. Usually this corresponds to keeping only the most important (highest weight) states in the expansion, checked through saturation of appropriate sum rules (see, e.g., Ref.~\cite{caux2009correlation}). However, for a judicious choice of initial state (such as~\eqref{Eq:InitialState}) that consists of a finite sum of eigenstates, Eq.~\eqref{Eq:ExpectationValue} becomes a finite sum of matrix elements and it is not necessary to truncate the sums -- evaluation of Eq.~\eqref{Eq:ExpectationValue} is, instead, exact. 

From Eq.~\eqref{Eq:ExpectationValue} we see that there is another important consideration in the computation of time-evolved expectation values: we must know expressions for matrix elements $ \langle \{p\};\{\lambda\}|\hat O(0) | \{k\}; \{\mu\} \rangle$ of the operator of interest $\hat O(0)$. In the case of the Yang-Gaudin model, matrix elements of certain operators have been derived from the algebraic Bethe ansatz in Refs.~\cite{pozsgay2012on,pakuliak2015form}. We note that determinant representations of matrix elements \textit{are crucial for efficient numerical evaluation of dynamics}, both in- and out-of-equilibrium. In contrast, direct computations from the coordinate Bethe ansatz are able to obtain expressions for matrix elements (see, for example, Ref.~\cite{zill2016coordinate} in the Lieb-Liniger model), but evaluating these scales in complexity as $(N!)^2$ and, as such, are limited to very small numbers of particles $N\le 7$.

\section{Non-equilibrium dynamics}
\label{Sec:NonEq}

We now consider the time-evolution of expectation values when starting from the initial state~\eqref{Eq:InitialState}. By construction, this initial state is a linear superposition of eigenstates with different energies and thus undergoes non-trivial time-evolution. The central object of our study will remain the expectation values of the density operators 
\begin{align}
  \rho_\sigma(x,t) = \big\langle \{I\}\big|e^{iHt} \Psi^\dagger_\sigma(x) \Psi_\sigma(x) e^{-iHt} \big| \{I\}\big\rangle.
\end{align}
We evaluate this via the matrix element expansion discussed in the previous section. Our focus here is on strongly interacting Yang-Gaudin Fermi and Bose gases, $\gamma \gg 1$. We briefly discuss weak coupling in the Fermi gas in \ref{App:Weakly}. 

\subsection{Light cone spreading of spin excitations in the Yang-Gaudin Fermi gas}
\label{Sec:LightConeFermi}

\begin{figure}[t]
\begin{center}
 \includegraphics[width=0.6\textwidth]{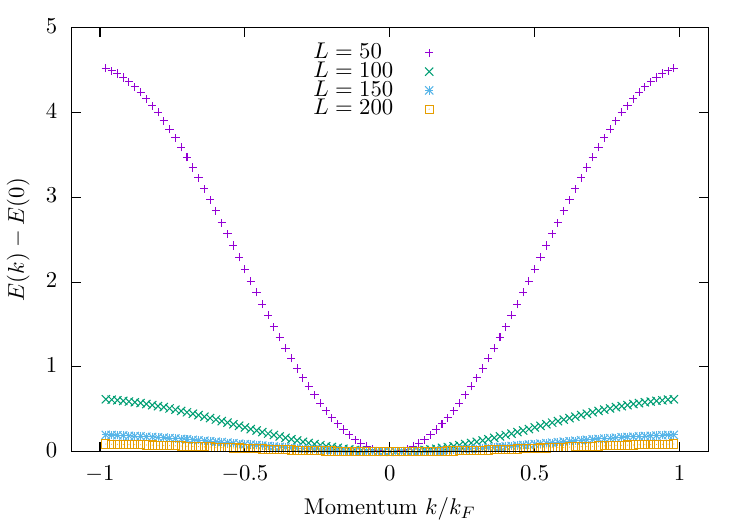}
\end{center}
\caption{The dispersion relation for the spin excitations contained in the initial state~\eqref{Eq:InitialState} for the Fermi Yang-Gaudin gas. Data is presented for $N=100$ particles with interaction parameter $c=50$ for a number of system sizes, as a function of momentum in units of the Fermi momentum $k_F = \pi \varrho$. Recall that $\gamma = c/\varrho$ and $\varrho = N/L$.}
\label{Fig:Dispersion_Fermions}
\end{figure}

Let us begin by developing some intuition. In Sec.~\ref{Sec:InitialState}, we offered an interpretation of the initial state~\eqref{Eq:InitialState} as a linear superposition of spin excitations. Each of these excitations has a momentum $P$ determined by the integers $\{I\}$ and $J$ entering the logarithmic Bethe ansatz equations. The dispersion for these spin excitations is presented in Fig.~\ref{Fig:Dispersion_Fermions}. In the strong coupling limit, $\gamma \gg 1$, the velocity of the spin excitations with momentum $P \ll \varrho$ are~\cite{mcguire1965interacting}
\begin{align}
v_s = 2P \bigg( \frac{1}{N} + \frac{2\pi^2}{3\gamma}\bigg) + O\Big( \gamma^{-2} \Big). 
\label{Eq:spinvelo}
\end{align}
For very large numbers of particles $N \gg \gamma/\pi^2$ the first term in the bracket is negligible, and the time $t_\ast$ it takes for a spin excitation of momentum $P$ to travel unit distance is 
\begin{align}
t_* = \frac{1}{v_s} \approx \frac{3\gamma}{4\pi^2 P}.
\label{Eq:tstar}
\end{align}

Equation~\eqref{Eq:tstar} allows us to develop two pieces of intuition. Firstly, it is clear that the light cone will be defined by `large' momentum $P$ spin excitations, which is not entirely unsurprising~\footnote{Indeed, from Fig.~\ref{Fig:Dispersion_Fermions} it is clear that modes with $P\sim k_F/2$ define the light cone.}. Secondly, if we construct a dimensionless parameter by rescaling $t_\ast$ by the dimensionless interaction parameter $\gamma$ and the Fermi time $t_F = 1/E_F = 1/(\varrho \pi)^2$, we find that $t_*/(\gamma t_F) = 3 \varrho^2/4P$ is an \textit{interaction-independent quantity}. Thus at fixed density $\varrho$, we expect an approximate collapse of data plotted as a function of $t/(\gamma t_F)$. 

\begin{figure}[t]
\begin{center}
\includegraphics[trim=170 75 78 165,clip,width=0.325\textwidth]{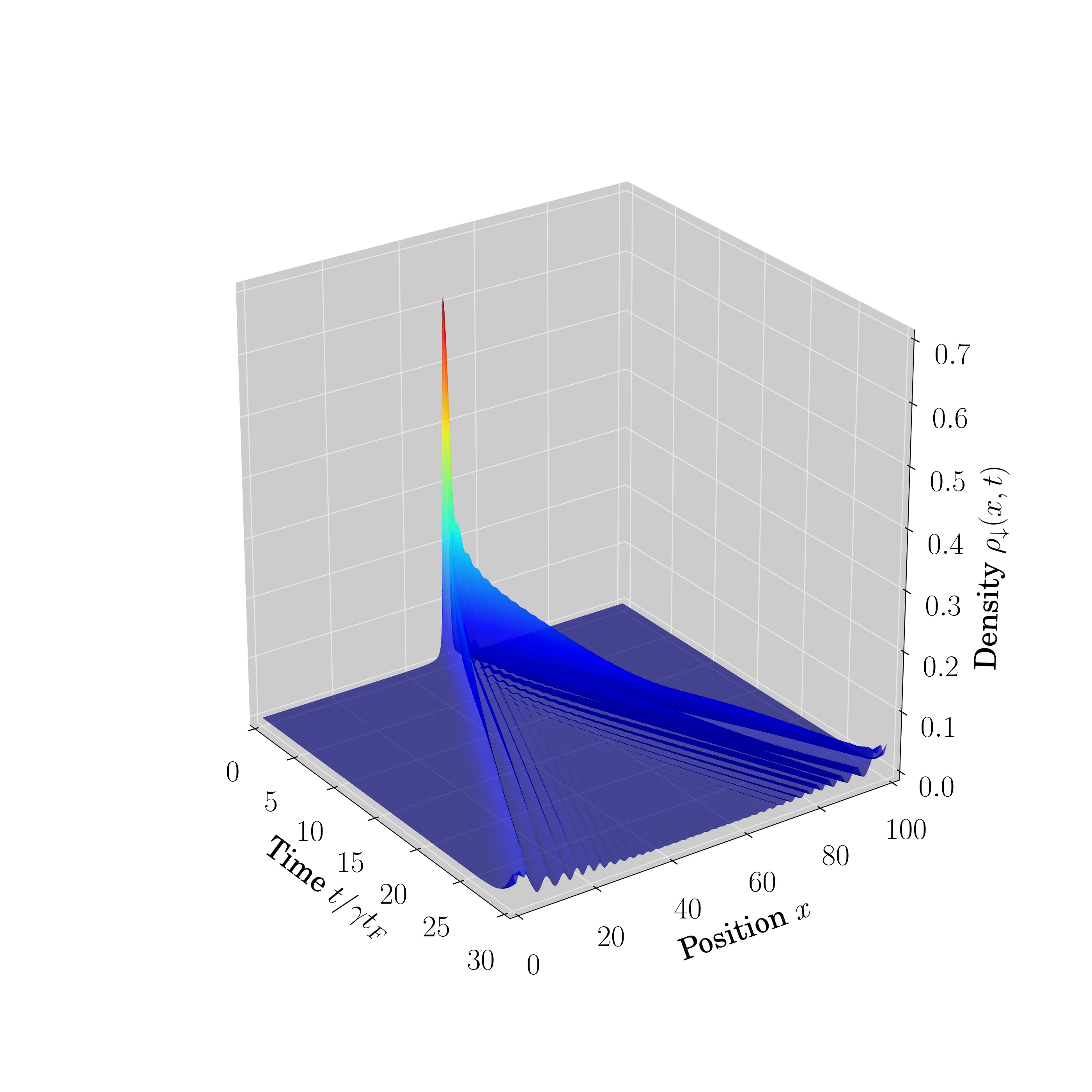} 
\includegraphics[trim=170 75 78 155,clip,width=0.325\textwidth]{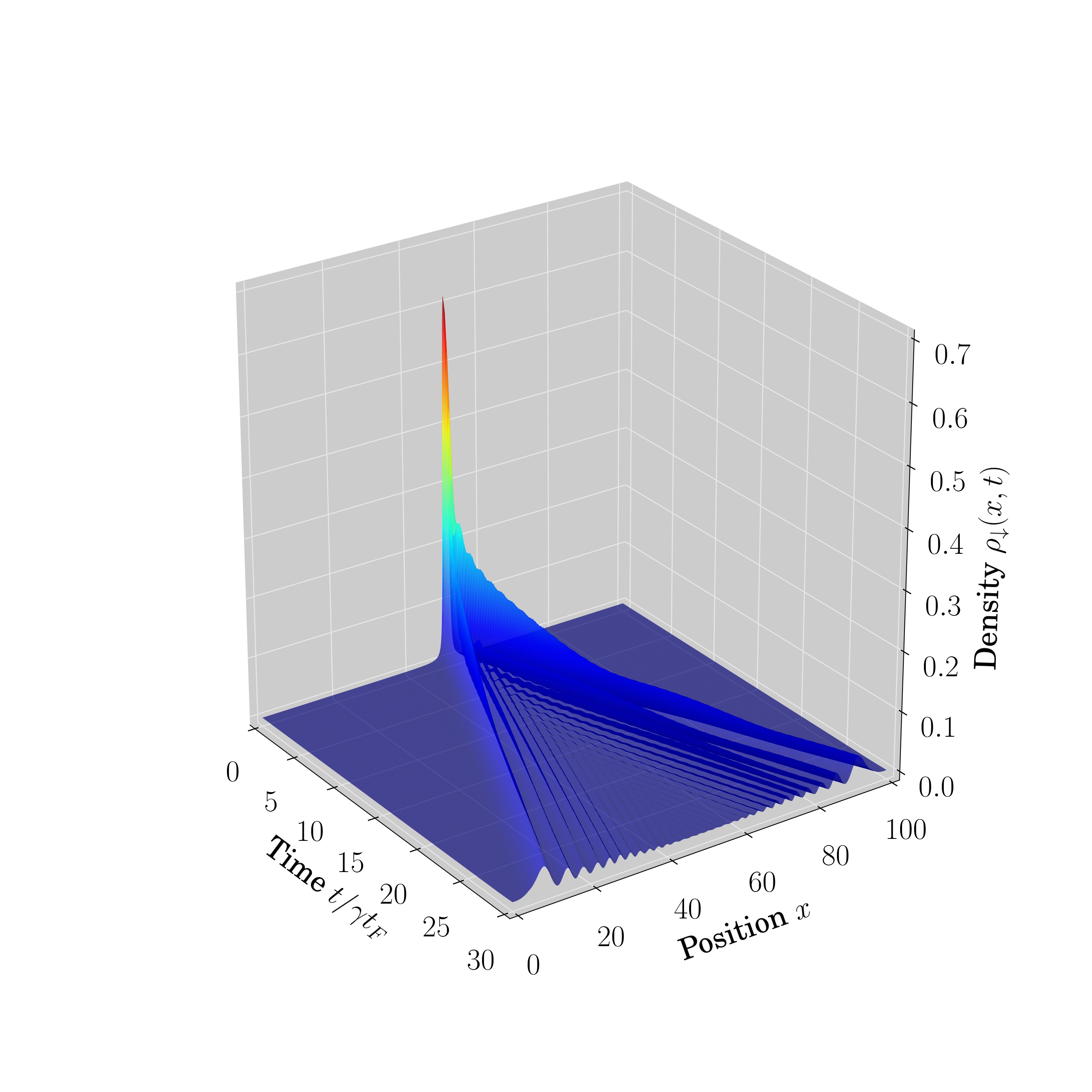} 
\includegraphics[trim=170 75 78 165,clip,width=0.325\textwidth]{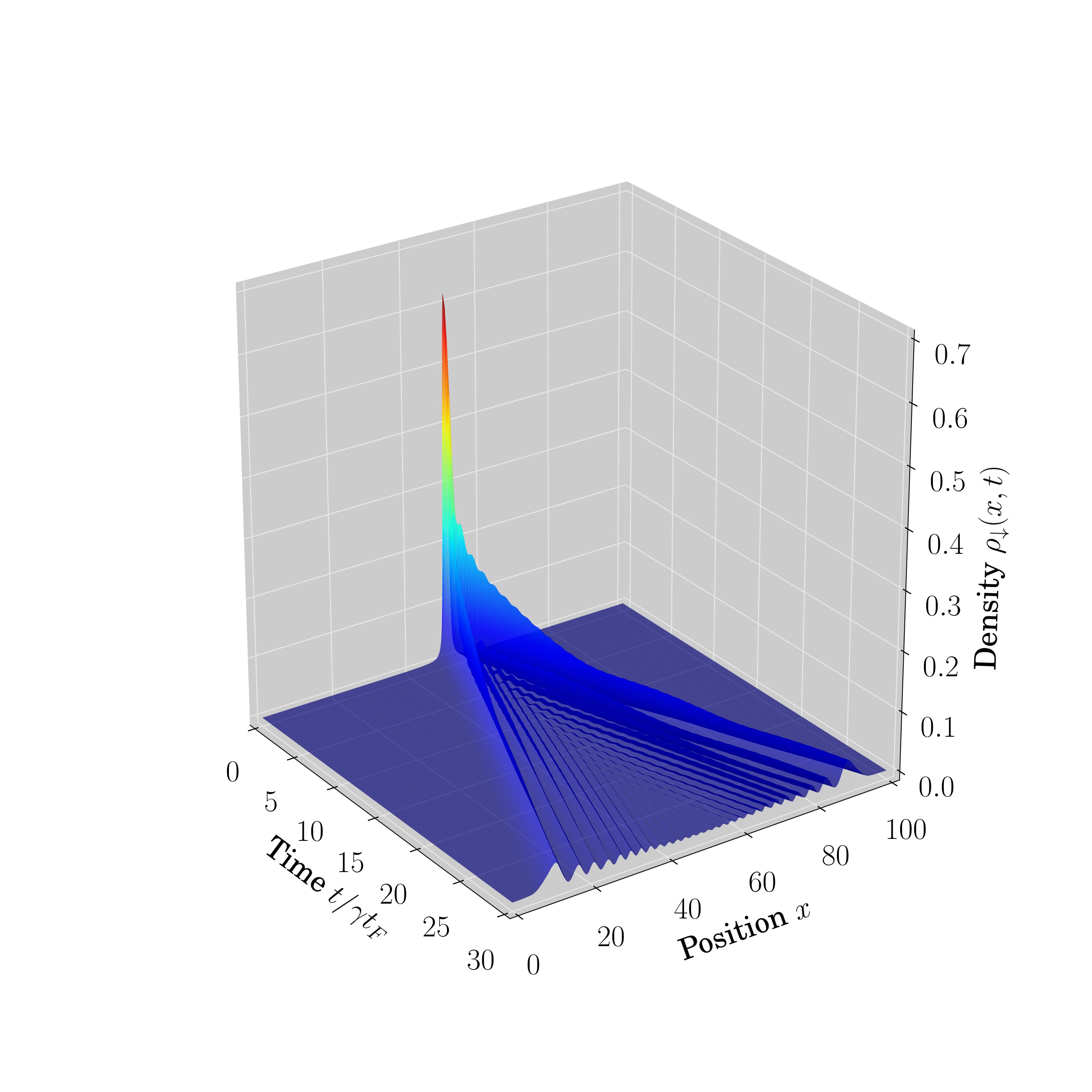}
\end{center}
\caption{Time-evolution of the minority species density profile, $\rho_\downarrow(x,t)$, in the Yang-Gaudin Fermi gas. We consider $N=100$ fermions on the length $L=100$ ring with interaction strength (left) $\gamma=100$; (center) $\gamma=50$; (right) $\gamma=25$. Data is plotted as a function of the rescaled time, $t/\gamma t_F$, and exhibits an approximate collapse.}
\label{Fig:Rescale}
\end{figure}

In Fig.~\ref{Fig:Rescale} we show an example of this approximate collapse upon rescaling time by $\gamma t_F$, where the time-evolution of the minority density $\rho_\downarrow(x,t)$ is shown for $\gamma = 25, 50, 100$. The scaling is evidenced through the similarity of the three figures. The data shows a clear light cone effect, with the maximal velocity governed by spin excitations with momentum $P \sim k_F/2$, see Fig.~\ref{Fig:Dispersion_Fermions}. Modes with lower velocities are observed as small oscillations within the light cone. A similar collapse upon rescaling is observed when $\gamma$ is varied at fixed interaction strength $c$ and number of particles $N$ through change of the system size $L$.   

\begin{figure}
\begin{center}
\includegraphics[trim=170 75 80 155,clip,width=0.5\textwidth]{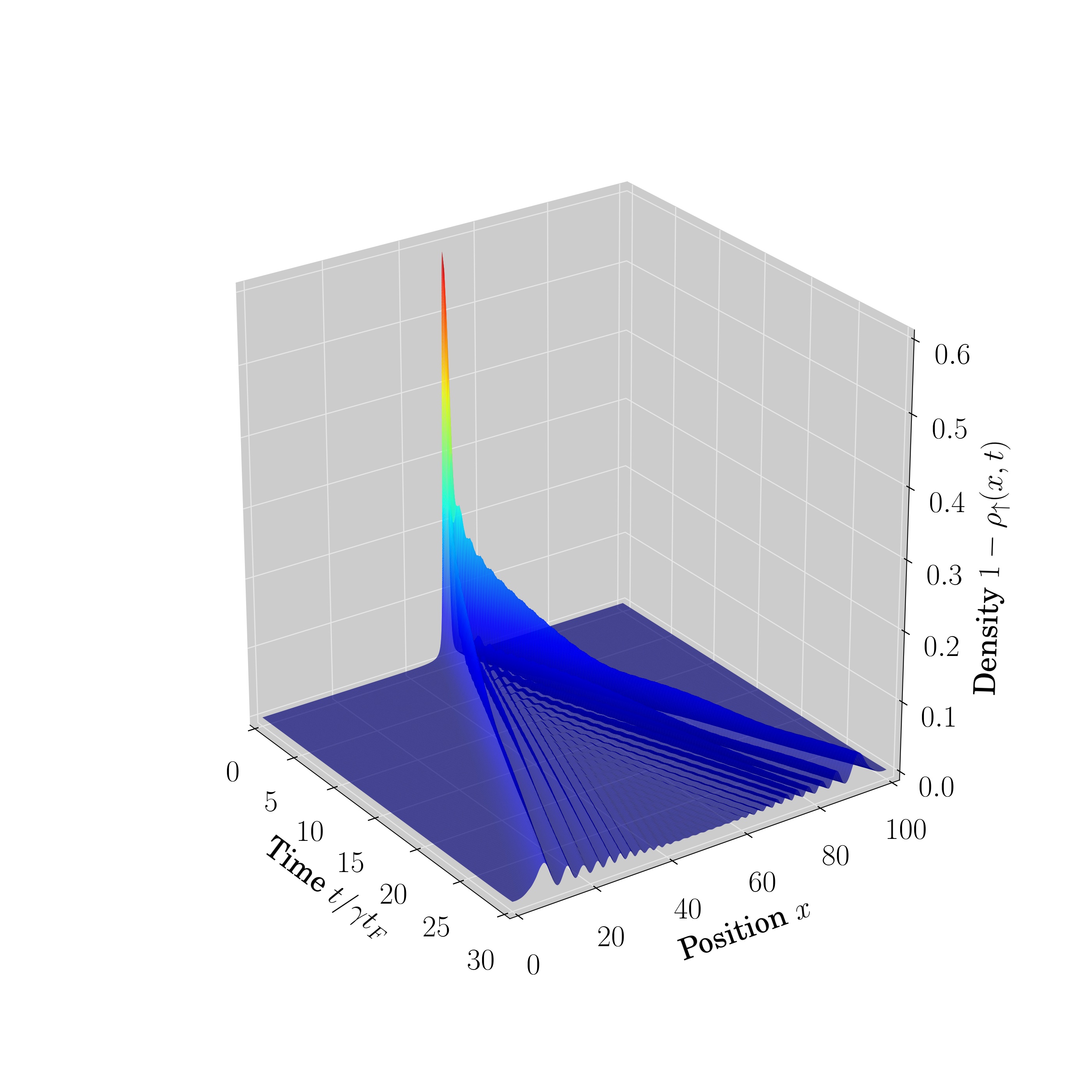}
\end{center}
\caption{Time-evolution of the density notch $1-\rho_\uparrow(x,t)$ for the initial state~\eqref{Eq:InitialState} in the Yang-Gaudin Fermi gas. We consider $N=100$ particles on the length $L=100$ ring with interaction parameter $\gamma = 50$. Data is plotted as a function of the rescaled time, $t/\gamma t_F$, and exhibits a similar scaling collapse upon varying $\gamma$ to Fig.~\ref{Fig:Rescale}.}
\label{Fig:FermionMajorityc50}
\end{figure}

The behavior of the majority species density, $\rho_\uparrow(x,t)$, is very closely related to that presented in Figs.~\ref{Fig:Rescale}. In Fig.~\ref{Fig:FermionMajorityc50} we present the time-evolution of the density notch, $1 - \rho_\uparrow(x,t)$, for parameters equivalent to Fig.~\ref{Fig:Rescale}(b). We see that the time-evolution of the density notch and the minority species are very similar, with only minor differences in amplitude.  We have worked here with $N=100$ particles, but we note that for the presented times we are, in effect, in the infinite volume limit. To further demonstrate this, we show an example simulation with $N=1000$ fermions in \ref{App:LargeN}.

\subsection{Light cone spreading of spin excitations in the Yang-Gaudin Bose gas}
Let us now turn our attention towards the nonequilibrium dynamics in the Yang-Gaudin Bose gas. We remind the reader that we study the minority component only; see the discussion of Sec.~\ref{Sec:Densities}. 

\subsubsection{Scaling collapse of data at strong coupling.}
\begin{figure}[t]
\begin{center}
\begin{tabular}{cc}
\includegraphics[trim=170 75 80 165,clip,width = 0.5\textwidth]{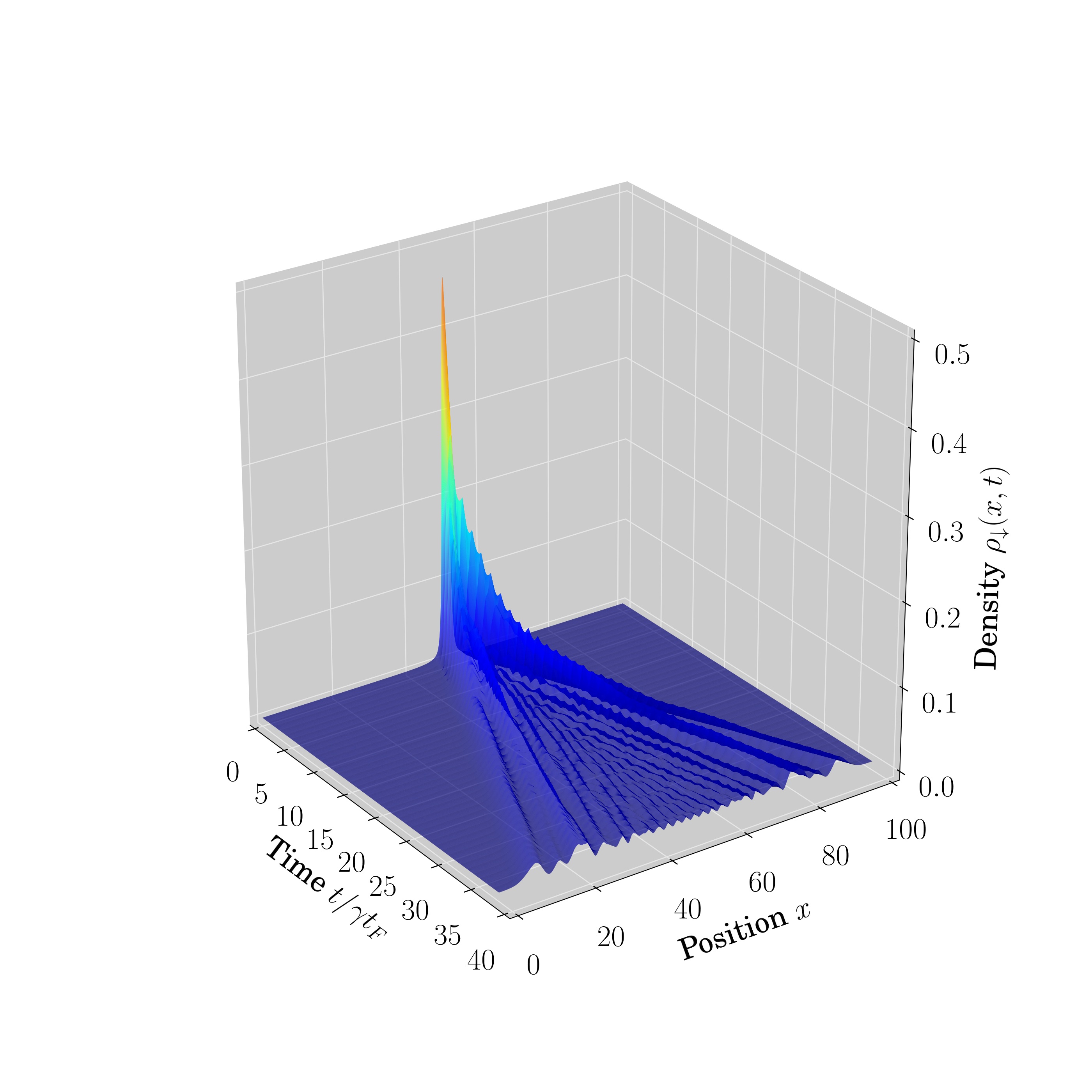} 
\includegraphics[trim=170 75 80 165,clip,width = 0.5\textwidth]{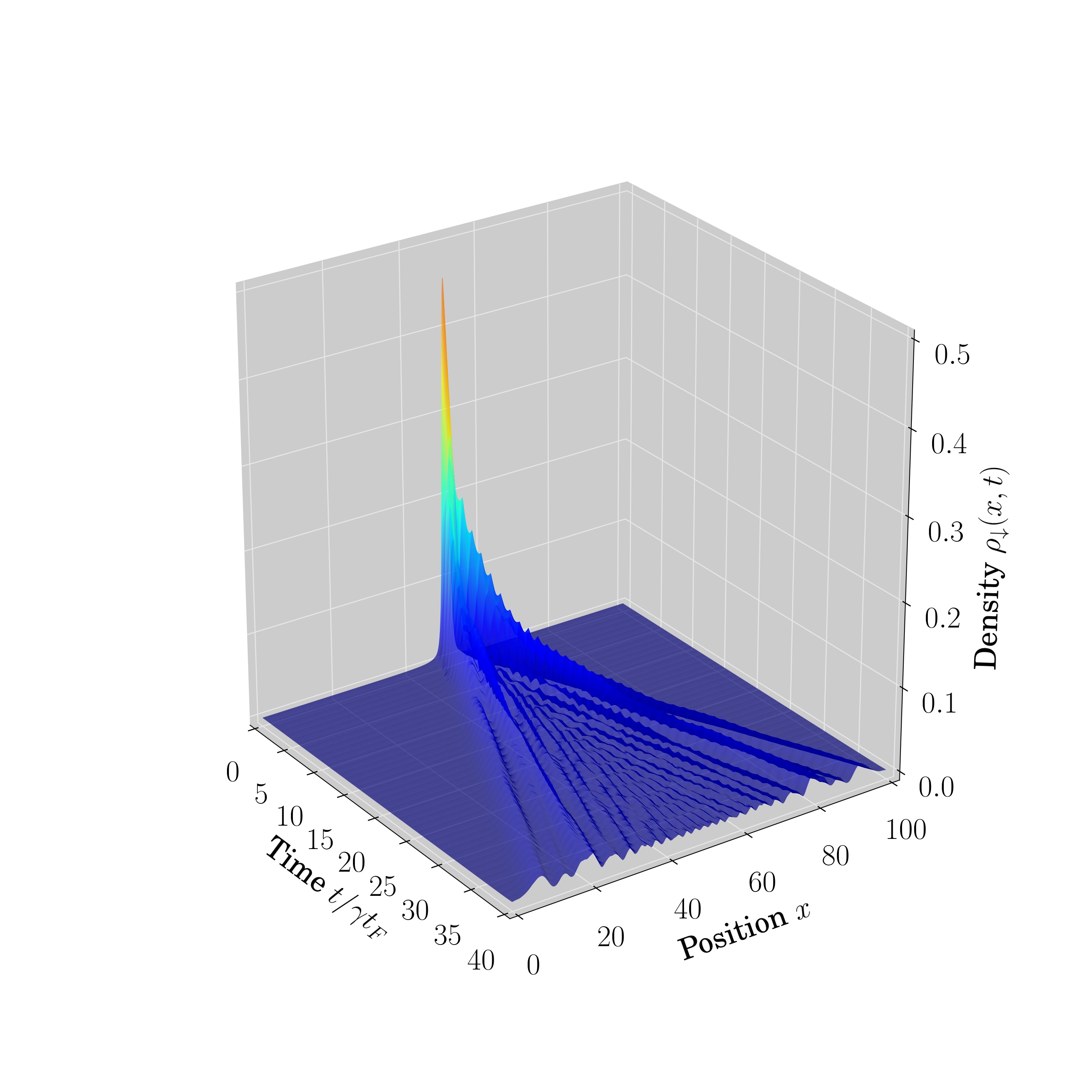} 
\end{tabular}
\end{center}
\caption{Time-evolution of the minority species density profile $\rho_\downarrow(x,t)$ for the initial state~\eqref{Eq:InitialState} in the Yang-Gaudin Bose gas. The state was constructed for $N = 100$ bosons at unit filling with dimensionless interaction parameter (left panel) $ \gamma = 50$; (right panel) $\gamma = 25$. Rescaling the time-axis to $t/\gamma t_F$ results in the two cases `collapsing' on to one another.}
\label{Fig:Boson_cdep}
\end{figure}

In Fig.~\ref{Fig:Boson_cdep}, we present the time-evolution of the minority species density, $\rho_\downarrow(x,t)$, for two values of the interaction strength $\gamma = 25,\, 50$. As with the Fermi gas, we see that the two cases collapse on top of one another upon rescaling of the time axis, $t/\gamma t_F$. The origin of this is identical to the case of the Fermi gas discussed in Sec.~\ref{Sec:LightConeFermi}: at strong coupling $\gamma \gg 1$ the expression for the spin velocity in the Yang-Gaudin Fermi and Bose gas coincides~\cite{fuchs2005spin,batchelor2006collective}, see Eq.~\eqref{Eq:spinvelo}. This is a natural consequence of the fermionization of the bosons at strong coupling: at $\gamma = \infty$ the bosons obey the local hardcore interaction constraint $\Psi^\dagger_\sigma(x) \Psi^\dagger_\sigma(x) = 0$, much like fermions. 

\begin{figure*}
\begin{center}
\begin{tabular}{ll}
(a) & (b) \\
\includegraphics[trim=170 75 80 165,clip,width = 0.45\textwidth]{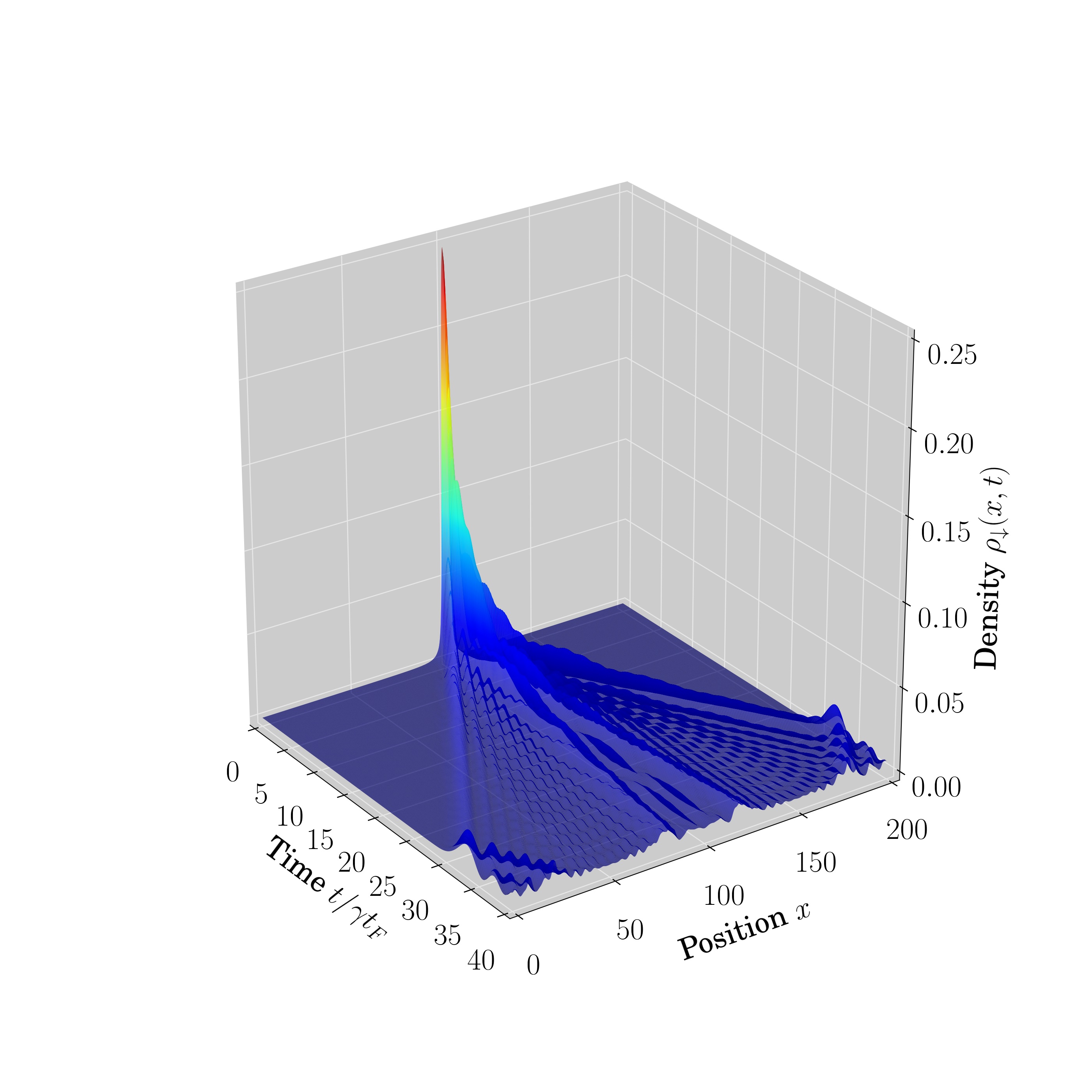} &  
 \includegraphics[trim=170 75 80 165,clip,width = 0.45\textwidth]{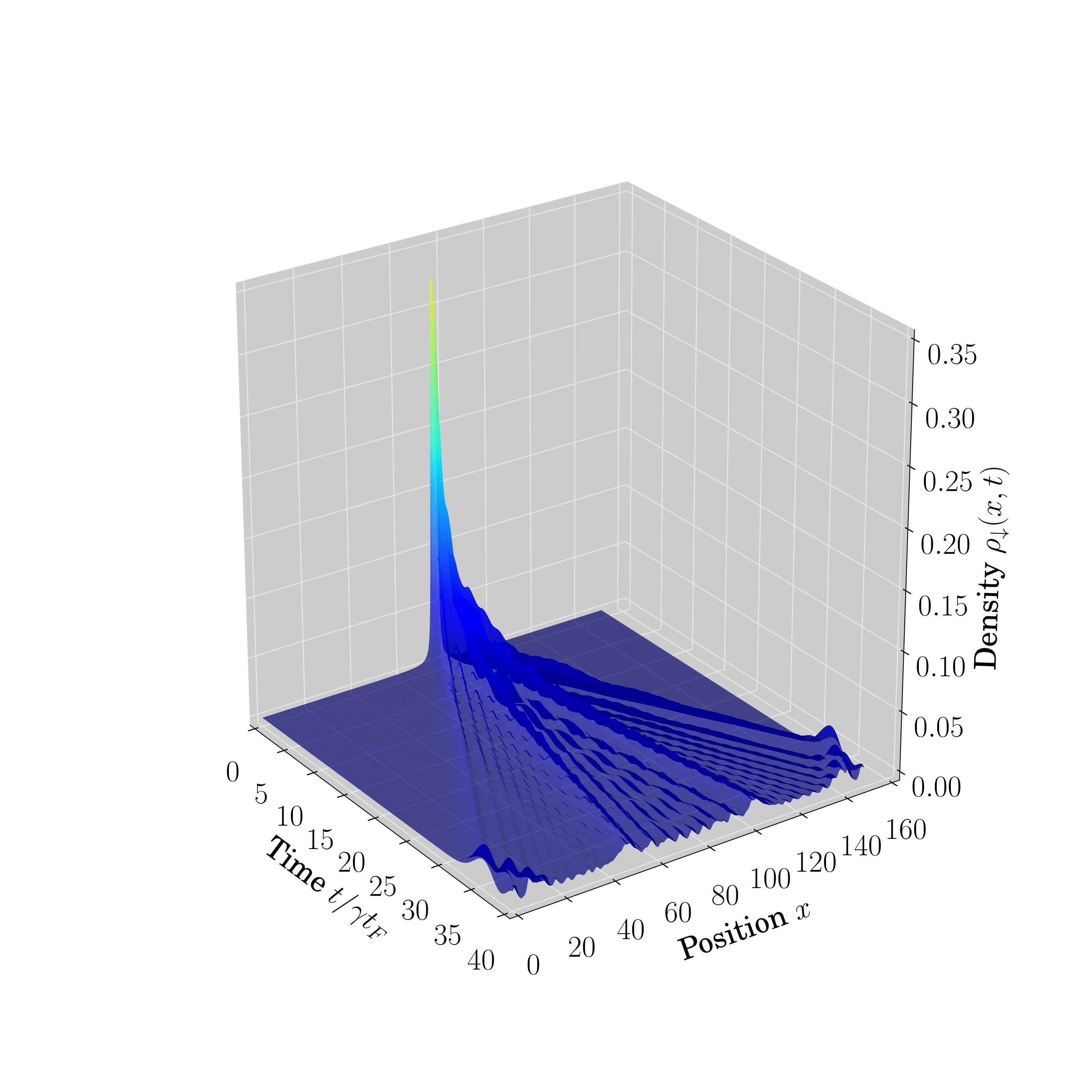} \\
(c) & (d)  \\
\includegraphics[trim=170 75 80 165,clip,width = 0.45\textwidth]{Bosons_c50_L100_N100_s}  &
\includegraphics[trim=170 75 80 165,clip,width = 0.45\textwidth]{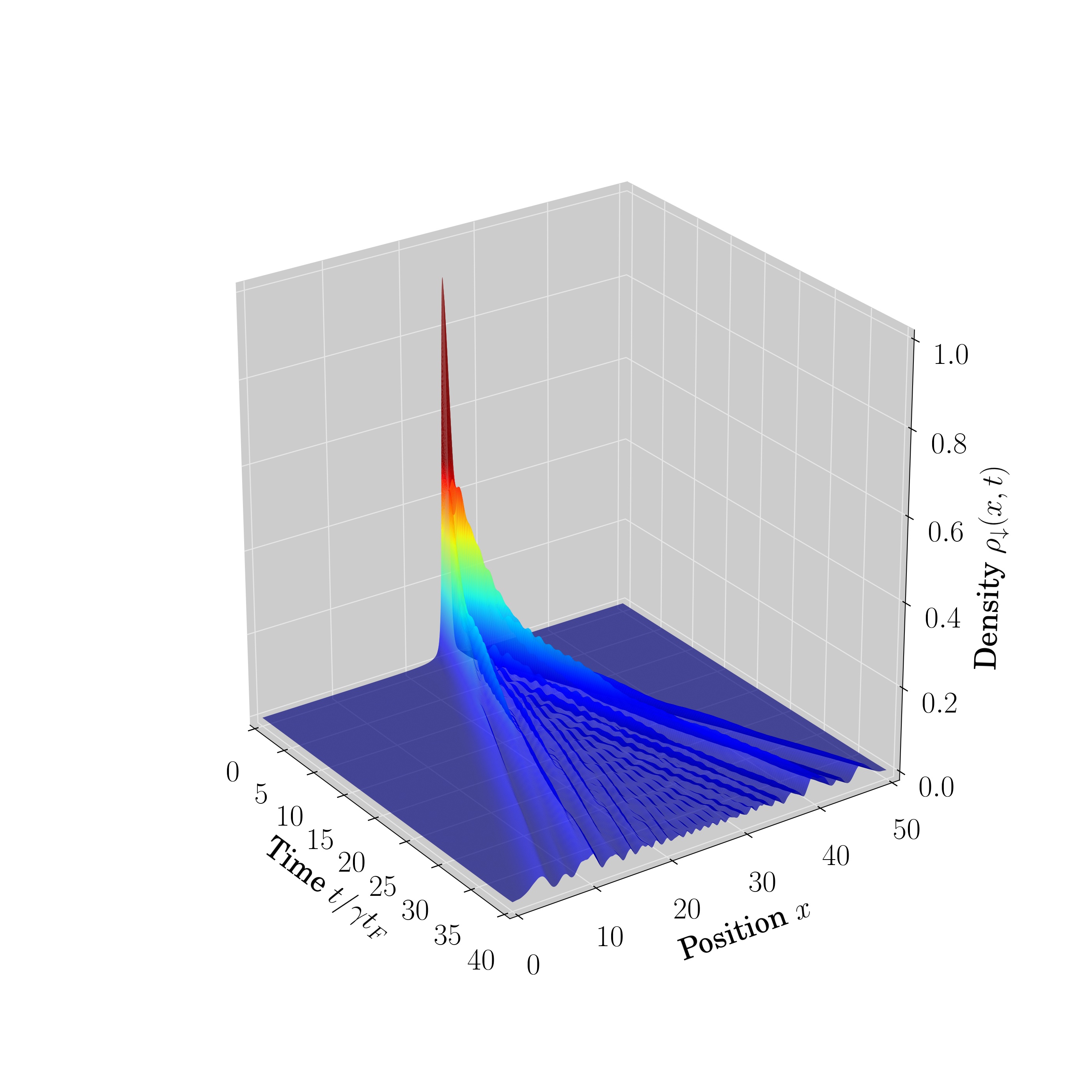}
\end{tabular}
\end{center}
\caption{The time-evolution of the density profile of the minority species $\rho_\downarrow(x,t)$ in the Yang-Gaudin Bose gas. The initial state~\eqref{Eq:InitialState} is prepared with $N=100$ particles with interaction strength $c=50$ for a number of system sizes $L$: (a) $L=200,\, \gamma=100$; (b) $L=150,\, \gamma=75$; (c) $L=100,\, \gamma=50$; (d) $L=50,\,\gamma=25$. The time axis is rescaled by $\gamma t_F$. A second light cone, propagating at a slower velocity, is clearly observed at lower densities.}
\label{Fig:Boson_Ldep}
\end{figure*}

\subsubsection{Multiple light cones.}

One interesting and immediately obvious difference between the Yang-Gaudin Fermi gas, Fig.~\ref{Fig:Rescale}, and the Bose gas, Fig.~\ref{Fig:Boson_cdep}, is the apparent ``messiness'' of the bosonic problem. The Fermi gas, Fig.~\ref{Fig:Rescale}, shows a single clean light cone with small sub-leading oscillations in the interior of the cone. On the other hand, the Bose gas appears to have multiple strong oscillations at close to the light cone velocity, Fig.~\ref{Fig:Boson_cdep}.

This situation is elucidated by computing the time-evolution for a variety of average densities $\varrho$, as shown in Fig.~\ref{Fig:Boson_Ldep}, where the number of particles $N=100$ and the interaction strength $c=50$ are fixed and the system size $L$ varied (hence the dimensionless interaction parameter $\gamma$ is varied). Indeed, we see that at lower densities (i.e., larger system sizes $L$ and larger $\gamma$) there are clearly \textit{two dominant contributions} to the non-equilibrium dynamics, leading to a \textit{double light cone structure}. The same is true at higher densities, although at the time scales plotted the two light cones are not widely separated. Interestingly, the slower light cone behaves very differently to the leading front: the excitations contributing to the second light cone propagate through the gas with a velocity that does not exhibit scaling collapse when time is scaled $t \to t/\gamma t_F$. 

The appearance of this second light cone, with behavior inconsistent with the leading edge is quite surprising, and is clearly physics different from that displayed in the Fermi gas. We will now spend some time discussing this. 

The first point to establish is: What is the  origin of the `slow mode' that provides a clear second light cone within the leading light cone? Let us first rule out two scenarios: (i) small momentum spin excitations $P \ll \varrho$ with velocity described by Eq.~\eqref{Eq:spinvelo}; (ii) charge excitations moving with Fermi velocity $v_F \approx 2\pi \varrho$. In both of these cases, it is easy to show that the time taken for excitations to propagate $L/2$ does not scale in the same manner with change in density $\varrho$ as that observed for the slow mode. 

\begin{figure}
\begin{center}
\includegraphics[width=0.6\textwidth]{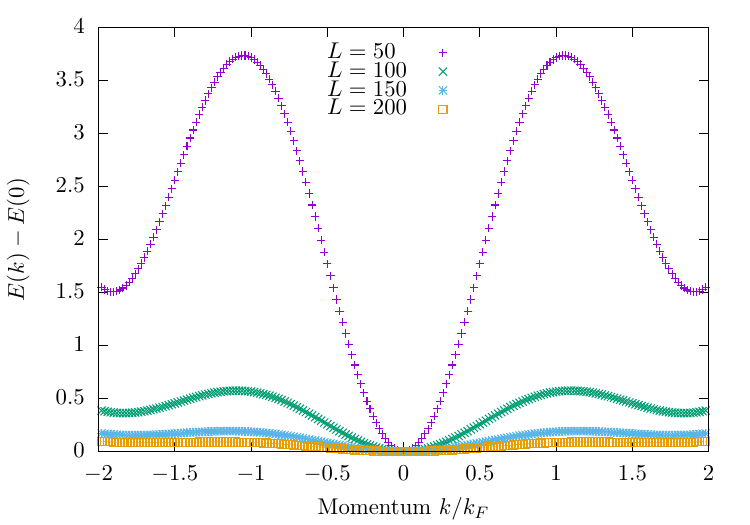}
\end{center}
\caption{The dispersion relation for the states in the initial state~\eqref{Eq:InitialState} for the Yang-Gaudin Bose gas. We construct the initial state with $N=100$ bosons and interaction parameter $c=50$ for a number of system sizes $L$. We rescale the momentum by the Fermi momentum $k_F = \pi \varrho$. For comparison, see Fig.~\ref{Fig:Dispersion_Fermions} for the analogous relation in the Fermi gas.}
\label{Fig:Dispersion_Bosons}
\end{figure}

Instead, let us turn our attention to higher momentum spin excitations -- is there some unusual behavior in the Bose gas that is not present in the Fermi gas (cf. Fig.~\ref{Fig:Dispersion_Fermions})? Indeed this is the case, as has recently been discussed in Ref.~\cite{robinson2017excitations}. The dispersion relation for spin excitations in the bosonic initial state~\eqref{Eq:InitialState} is shown in Fig.~\ref{Fig:Dispersion_Bosons}, where we see that this is surprisingly different to the Fermi gas, Fig.~\ref{Fig:Dispersion_Fermions}. 

One immediate difference is the range of momentum of the spin excitations in the Bose gas. We see that spin excitations in the Bose gas have momentum $-2k_F \le P \le 2k_F$, compared to $-k_F \le P \le k_F$ in the Fermi gas. This is direct consequence of the selection rules~\eqref{Eq:QNs1} for the integers $I_j$ in the Bose gas: in the Fermi sea configuration this leads to $\big| \sum_j I_j \big| = N/2$, which when combined with the bounds of the spin integer $|J| \le N/2$ naturally leads to $|P| \le 2k_F$. 

If we examine Fig.~\ref{Fig:Dispersion_Bosons} and restrict our attention to $|P|\le k_F$, we see that the spin excitations in this region are rather reminiscent of those observed in the Fermi gas, Fig.~\ref{Fig:Dispersion_Fermions}. This strongly suggests that the new physics in the Bose gas is related to spin excitations with momentum $|P| > k_F$. In this region, we see a rather interesting feature: a finite momentum minima in the energy, much like a roton excitation~\cite{landau1941theory,landau1947theory,feynman1954atomic,cohen1957theory}.  Close to momentum $P_r \sim 2k_F$ there are \textit{gapped spin excitations} with dispersion~\cite{robinson2017excitations} ($|P-P_r| \ll \varrho$)
\begin{align}
E_r(P) \approx \Delta_r + \frac{(P-P_r)^2}{2m_r}, 
\label{Eq:roton}
\end{align}
where $m_r$ is the effective mass of these gapped `roton-like' excitations and $\Delta_r$ is the energy gap.

\begin{figure}
\begin{center}
\includegraphics[width=0.6\textwidth]{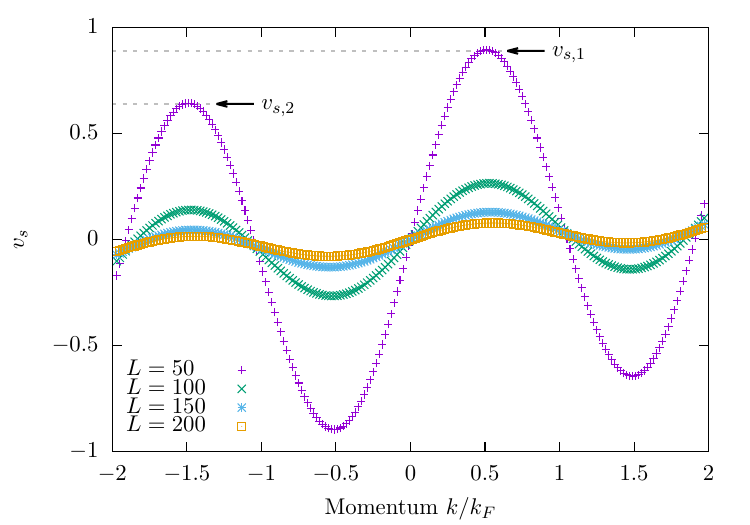}
\end{center}
\caption{The velocity $v_s$ of spin excitations with momentum $k$ in the Yang-Gaudin Bose gas. We consider a system of $N=100$ particles with interacting strength $c=50$ for a number of system sizes $L$. The momentum is plotted in units of the Fermi momentum $k_F = \pi \varrho$. We explicitly highlight the peak velocities $v_{s,1}$, $v_{s,2}$ discussed in the text.}
\label{Fig:spinvelo}
\end{figure}

The behavior of the inner light cone can be understood through the behavior of the effective spin velocity for the excitations with $|P| > k_F$. We define the effective spin velocity $v_s$ through  
\begin{align}
v_s = \frac{{\rm d} E_J}{{\rm d} P_J},
\end{align}
where $E_J$, $P_J$ are the energy and momentum of the state with spin rapidity integer $J$ in the initial state~\eqref{Eq:InitialState}. We present this in Fig.~\ref{Fig:spinvelo} for a number of different system sizes at fixed particle number $N=100$ and interaction strength $c=50$. We see that there are two significant propagation velocities in the system, $v_{s,1}$ and $v_{s,2}$, corresponding to states with momentum close to $|P| \sim k_F/2$ and $|P| \sim 3k_F/2$. The ratio $v_{s,2}/v_{s,1}$ is not invariant upon changing the system size (and hence $\gamma$ for fixed $N$ and $c$): for the reported data
\begin{align}
\left. \frac{v_{s,2}}{v_{s,1}} \right|_{L=50} \approx 0.7, \qquad
\left. \frac{v_{s,2}}{v_{s,1}} \right|_{L=200} \approx 0.2. 
\end{align}
This is consistent with the behavior of the slow light cone seen in Fig.~\ref{Fig:Boson_Ldep}: the slow mode becomes slower compared to the leading light cone with increasing system size. Thus the spin excitations about the roton-like minima with momentum $|P|\sim 3k_F/2$ are responsible for the second light cone observed in the time-evolution of the bosonic initial state~\eqref{Eq:InitialState}. 

Like with the fermions, we have run a number of simulations with large numbers of bosons (here $N=500)$.  Reported in \ref{App:LargeN}, we show an example showing a similar double light cone phenomena for the bosonic gas after the gas has evolved in time.

\section{Discussion and conclusion}
\label{Sec:Conc}

In this work, we used the integrability of the Yang-Gaudin Fermi and Bose gases to investigate the nonequilibrium dynamics emerging from an initial state containing a localized wave packet of spin excitations. The simple form of the initial state~\eqref{Eq:InitialState} was heavily motivated by the work of Sato~{\it et al.}~\cite{sato2012exact} and it allows us to compute the time-evolution in a numerically exact manner for large numbers of interacting particles. Nonequilibrium light cone effects in non-relativistic free field theories have been studied in Ref.~\cite{bertini2017approximate}; here we have considered a strongly interacting non-relativistic case.

Under time-evolution the initially localized packet of spin excitations dissolves, with excitations propagating through the gas within a light cone. At fixed particle density $\varrho$, the light cone exhibits a scaling collapse upon $t \to t/\gamma t_F$. This can be understood from the strong coupling expansion of the spin excitation velocity $v_s$, Eq.~\eqref{Eq:spinvelo}. This scaling collapse occurs in both the Yang-Gaudin Fermi and Bose gases. 

An important differences between the nonequilibrium dynamics of the Bose and Fermi gases is the appearance of a secondary light cone in the Bose gas. This `slow mode' is not associated with $P\sim0$ spin excitations, nor charge excitations, but instead arises from large momentum spin excitations. Unlike the Fermi gas, the spectrum of spin excitations above the Fermi sea of integers $\{I\}^{(b)}$, see Eqs.~\eqref{Eq:Ib1} and~\eqref{Eq:Ib2}, extends to momentum $|P| < 2k_F$. In the extended region $|P|>k_F$, the spin excitation dispersion exhibits a pronounced roton-like minima, see Fig.~\ref{Fig:Dispersion_Bosons} (and Ref.~\cite{robinson2017excitations} for further details). Thus there are \textit{gapped quadratically-dispersing spin excitations}, Eq.~\eqref{Eq:roton}, present within the state. By computing the effective spin excitation velocity across the whole range of momenta, see Fig.~\ref{Fig:spinvelo}, we identify the spin excitations responsible for the two light cones: the leading light cone is associated with spin excitations of momentum $|P| \sim k_F/2$, whilst the secondary light cone arises from spin excitations with momentum $|P| \sim 3k_F/2$.  The latter excitations can be pictured as excitations about the roton-like minima. 

We want to stress that the origin of the multiple light cones observed in the Bose gas quenches we consider is the enhanced density of states coming from multiple quadratic minima in the one-dimensional spin wave dispersion relation. This is in contrast to, say, the multiple light cone structure that appears in quenches of multi-component spin chains, as observed in Ref.~\cite{modak2019nested}. There, it appears that the additional structure arises from distinct branches of excitations. It should also be contrasted to the double light cone structure observed in two decoupled Luttinger liquids with small population imbalances~\cite{langen2018}. Here the double light cone structure arises because each Luttinger liquids possesses excitations with distinct velocities, $v_i,~i=1,2$. In the case considered in Ref.~\cite{langen2018}, the initial state of interest (which arises from splitting a single Luttinger liquid into two) leads to correlators in the symmetric/antisymmetric channel depending on velocities symmetric/antisymmetric combinations of the velocities, $v_1\pm v_2$. Thus multiple light cones in correlation functions can be realised via a number of mechanisms.

One question we did investigate preliminarily is whether our initial state, Eq.~\eqref{Eq:InitialState}, possesses a significant overlap with a ``more physical'' state?  This is a natural question to ask as our initial state is constructed as a linear superposition of spin excitations at different momenta.  Given that this superposition does lead to a localized minority state, can the linear superposition indeed be thought of as Gaussian localized particle in disguise?  We treat this issue in \ref{App:spinflip}.  We find in general that the overlap is ${\cal O}(1)$ for reasonable numbers of particles, with the overlap improving with increasing interaction strength. Furthermore the overlaps are higher in the fermionic gas.

Our results suggest that gapped quadratically dispersing single-particle excitations can have clear signatures in nonequilibrium dynamics. In the scenario that we consider, the integrability of the Yang-Gaudin model means that spin excitations at all energies are stable, possessing infinite life times. It would be interesting to consider a case with weak integrability breaking: this opens the possibility for spin excitations with energy $E > 2\Delta_r$ to decay into multiple roton-like excitations. Such instabilities have been observed in high-density liquid helium~\cite{donnelly1981specific} and leave clear signatures in the dynamical structure factor, in the form of the Pitaevskii plateau~\cite{pitaevskii1959properties}. It may be possible to compute the dynamics in such a scenario with truncated spectrum methods (see, e.g., Refs.~\cite{james2017nonperturbative} and~\cite{rakovsky2016hamiltonian}).

\ack We thank Bruno Bertini for interesting and useful discussions surrounding this work.  We would also like to thank the Isaac Newton Institute, University of Cambridge, for hospitality during parts of this work. This work was supported by the Condensed Matter Physics and Materials Science Division, Brookhaven National Laboratory, in turn funded by the U.S. Department of Energy, Office of Basic Energy Sciences, under Contract DE-SC0012704 (NJR, RMK), and the FOM and NWO foundations of the Netherlands (JSC). NJR also received funding from the European Union's Horizon 2020 research and innovation programme under grant agreement No 745944. 

\appendix

\section{Localized spin excitations above the diluted Fermi sea}
\label{App:diluted}

In this appendix we consider the initial state~\eqref{Eq:InitialState} with a `diluted Fermi sea' of integers, $\{I\}_d$, as defined below and the subsequent nonequilibrium dynamics. This is a brief example of other possibilities within this class of states.

\subsection*{Appendix A.1. The diluted Fermi sea in the Yang-Gaudin Fermi Gas}

Let us construct the initial state~\eqref{Eq:InitialState} in the Yang-Gaudin Fermi gas with a different choice of the integers $\{I\}$, that we call the `diluted Fermi sea' of integers. Symmetrically about the origin, we occupy every-other-integer, as depicted below for $N=4$ particles
\begin{center}
        \halfintegers
        \border
        \intrange{-4}{5}
        \fillI{-3}{4}
        \holeadd{-2}{}
        \holeadd{0}{}
        \holeadd{1}{}
        \holeadd{3}{}
        \intLabel{ \left\{I\right\}_d^{(f)}}
        \render
\end{center}
That is, 
\begin{align}
\{ I \}^{(f)}_d = \left\{ - \frac{2N - 1}{2}, -\frac{2N-1}{2} + 2 , \ldots \right\}, 
\label{Eq:Id}
\end{align}
where symmetry about the origin is implied. 

\begin{figure}
\begin{center}
\includegraphics[trim=0 0 0 30,clip,width=0.48\textwidth]{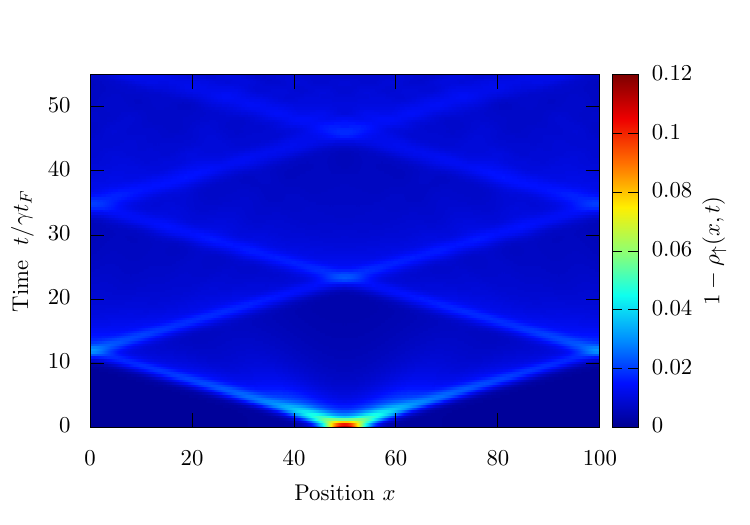} \includegraphics[trim=0 0 0 30,clip,width=0.48\textwidth]{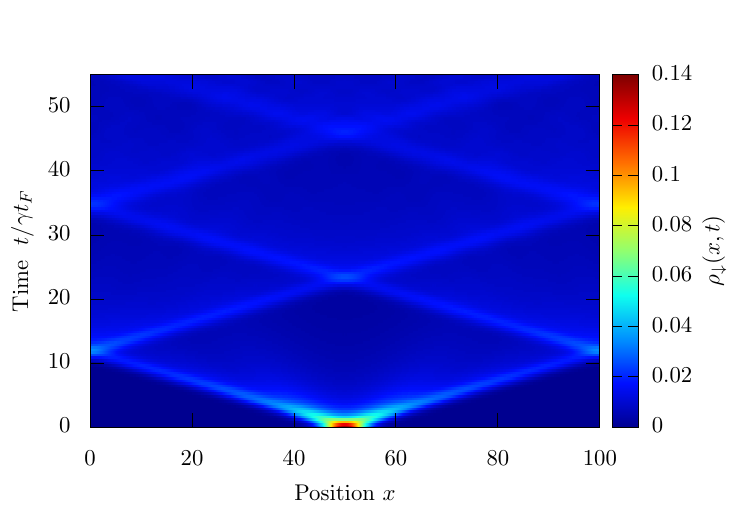}
\end{center}
\caption{The time-evolution of the density profile of the (left panel) majority $\sigma=\,\uparrow$; (right panel) minority $\sigma=\,\downarrow$ species  from the initial state~\eqref{Eq:InitialState} with the diluted Fermi sea of integers~\eqref{Eq:Id} in the Yang-Gaudin Fermi gas. The state is constructed for $N=100$ fermions at unit filling with interaction strength $c=10$.}
\label{Fig:NoneqId}
\end{figure}

The initial state~\eqref{Eq:InitialState} that we construct is now rather complicated, containing both spin and charge excitations (recall that charge excitations correspond to the particle-hole excitations about the filled Fermi sea of integers). Despite this change, the density profile of the initial state is quite similar (although broader) to that discussed in the main body of the paper. In Fig.~\ref{Fig:NoneqId} we present density plots of the nonequilibrium dynamics emerging from this state for $N=100$ particles at unit average filling with interaction parameter $c=10$. We see that the initially localized density dissolves, with spin excitations propagating in a light cone at almost twice the speed as those in the Fermi sea of integers. This is quite natural, as the momentum of the spin excitations should be approximately doubled in the diluted Fermi sea, leading to an approximately doubling of their velocities. 

\subsection*{Appendix A.2. The diluted Fermi sea in the Yang-Gaudin Bose gas}

\begin{figure}[t]
\begin{center}
\includegraphics[width = 0.6\textwidth]{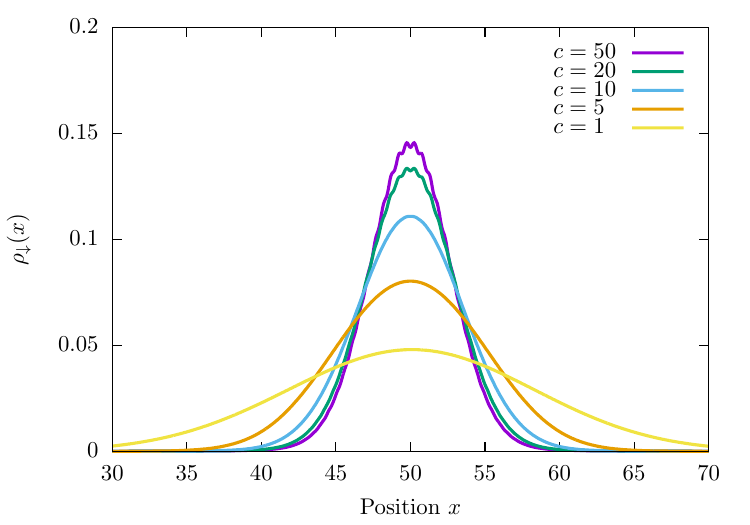}
\end{center}
\caption{The density $\rho_\downarrow(x)$ of the minority species density in the initial state~\eqref{Eq:InitialState} with the diluted Fermi sea of integers~\eqref{Eq:Id_bosons} in the Yang-Gaudin Bose gas. The state is constructed with $N=100$ bosons at unit filling for a number of interaction strengths.}
\label{Fig:Bosons_Diluted}
\end{figure}

\begin{figure}
\begin{center}
\includegraphics[trim=0 0 0 30,clip,width=0.6\textwidth]{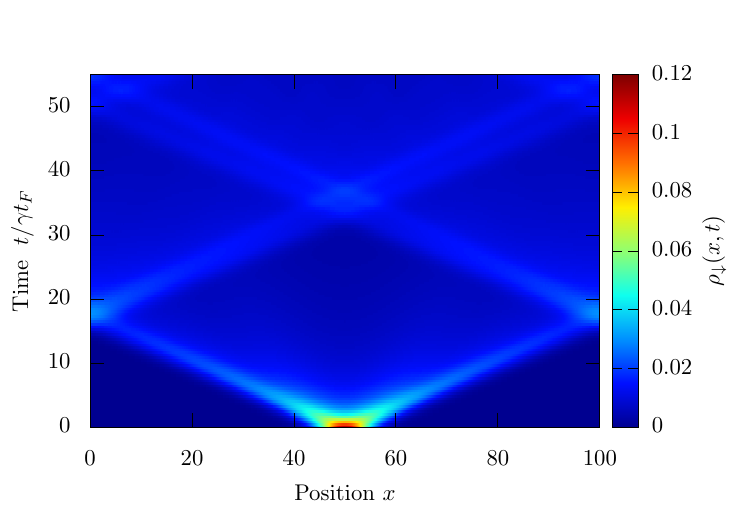}
\end{center}
\caption{The time-evolution of the density $\rho_\downarrow(x,t)$ in the Yang-Gaudin Bose gas starting from the initial state~\eqref{Eq:InitialState} with the diluted Fermi sea of integers~\eqref{Eq:Id_bosons}. The state is constructed with $N=100$ bosons at unit average density and interaction strength $c=10$.}
\label{Fig:NoneqId_bosons}
\end{figure}

Let us now consider the analogous state in the Yang-Gaudin Bose gas
\begin{align}
\{ I \}^{(b)}_d = \{ - N, -N + 2, \ldots \}.
\label{Eq:Id_bosons}
\end{align}
The initial density configuration is shown in Fig.~\ref{Fig:Bosons_Diluted}; we see that for similar interaction strength, the localization of the minority species is suppressed in comparison to the Fermi sea of integers, cf. Fig.~\ref{Fig:InitialDensityBosons}. Nevertheless, we still have a reasonably localized excitation. In Fig.~\ref{Fig:NoneqId_bosons} we present the nonequilibrium dynamics. Here we see that, once again, there is a double light cone structure clearly visible after times $t/\gamma t_F \sim 30$. This is reminiscent of the behavior observed in the initial state when constructed with the Fermi sea of integers, which arises due to the presence of roton-like spin excitations.

\section{Localized spin excitations above the shifted Fermi sea}
\label{App:shifted}

We now consider a second alternative example for the initial state~\eqref{Eq:InitialState}. This time we construct the state with the shifted Fermi sea of integers, $\{I\}_s$, as defined below. 

\subsection*{Appendix B.1. The shifted Fermi sea in the Yang-Gaudin Fermi gas}

In the Yang-Gaudin Fermi gas we construct the initial state~\eqref{Eq:InitialState} with a Fermi sea of integers that is no longer symmetric about the origin (i.e., shifted). We can, for example, consider the case where we only allow negative integers:
\begin{align}
\{ I \}^{(f)}_s = \left\{ - \frac{2N - 1}{2}, -\frac{2N-1}{2} + 1 , \ldots \right\}.
\label{Eq:Is}
\end{align}
For $N=6$ particles this corresponds to:
\begin{center}
        \halfintegers
        \border
        \intrange{-6}{2}
        \fillI{-5}{0}
        \intLabel{ \left\{I\right\}_s^{(f)}}
        \render
\end{center}
The initial state once again contains a well-localized wave packet of spin excitations. However, we note that constructing the initial state~\eqref{Eq:InitialState} with this configuration of integers leads to the state having a non-zero average momentum. This imprints itself on the state via a breaking of spatial inversion (about $L/2$) symmetry and has obvious signatures in the dynamics. 

\begin{figure}
\begin{center}
\includegraphics[trim=0 0 0 30,clip,width=0.45\textwidth]{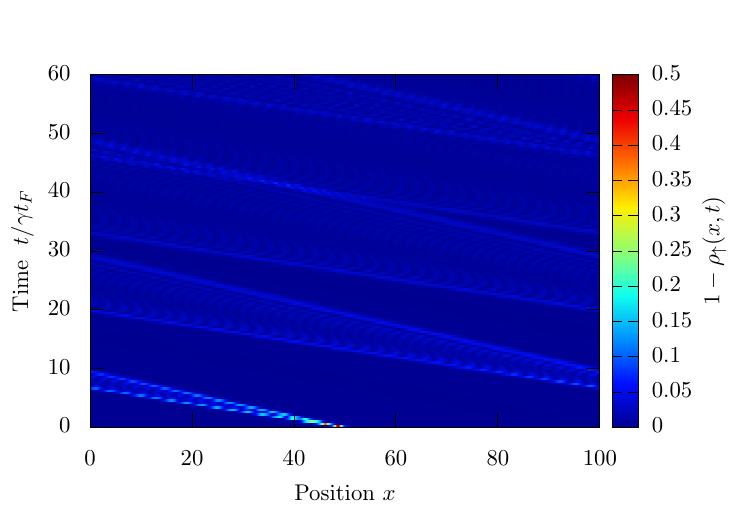}
\includegraphics[trim=0 0 0 30,clip,width=0.45\textwidth]{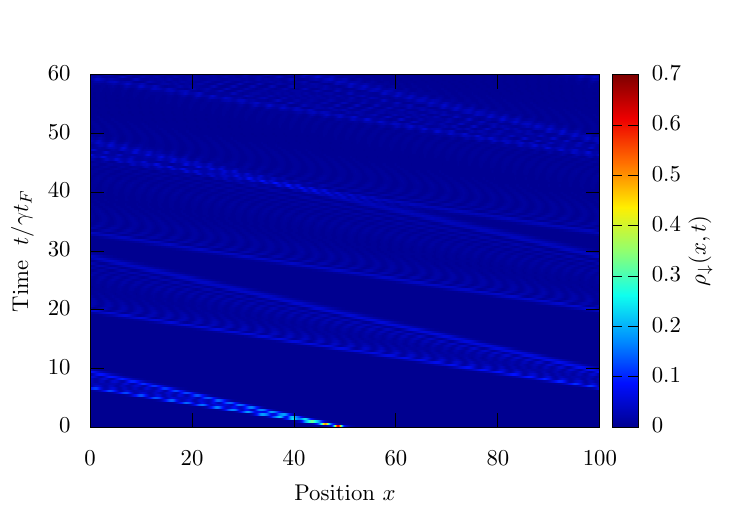}
\end{center}
\caption{The time-evolution of the density (a) $1-\rho_\uparrow(x,t)$; (b) $\rho_\downarrow(x,t)$ in the Yang-Gaudin Fermi gas starting from the initial state~\eqref{Eq:InitialState} with the shifted Fermi sea of integers~\eqref{Eq:Is}. The state is constructed for $N=100$ fermions at unit average density with interaction strength $c=10$.}
\label{Fig:NoneqIs}
\end{figure}

We present the nonequilibrium dynamics of the initial state with the shifted integers~\eqref{Eq:InitialState} in Fig.~\ref{Fig:NoneqIs}. Under time-evolution, the well-localized wave packet of spin excitations moves rapidly to the left, slowly delocalizing in the process.  After relatively short times, $t \sim 10t_F$, the wave packet has traversed the ring -- under continued time-evolution this repeats, with the wave packet becoming increasingly delocalized. The spreading of the wave packet is suppressed in comparison to the Fermi sea configuration of integers, which can be understood in terms of the spin excitation dispersion, cf. Fig.~\ref{Fig:Dispersion_Fermions}: the spreading of the wave packet is governed by the differences in velocity between the fastest and slowest \emph{left-moving spin excitation}. For the Fermi sea configuration, instead the spreading is governed by the \textit{difference in velocity between the fastest left- and fastest right-moving spin excitations}. 

\subsection*{Appendix B.2. The shifted Fermi sea in the Yang-Gaudin Bose gas}

\begin{figure}[t]
\begin{center}
\includegraphics[width = 0.6\textwidth]{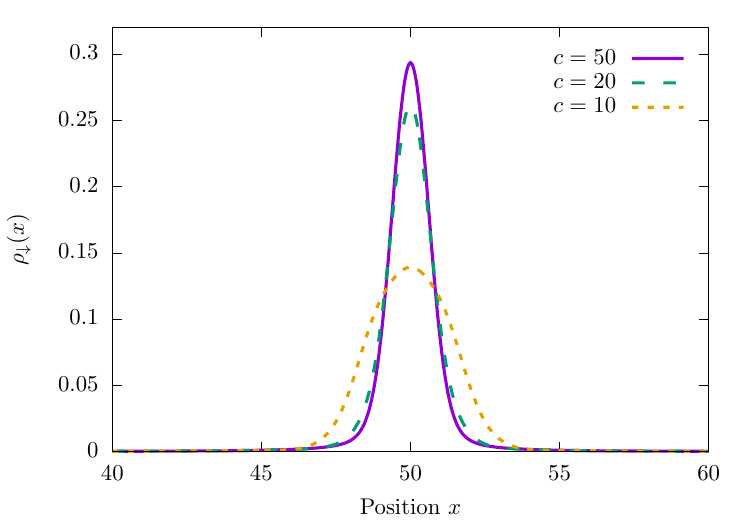}
\end{center}
\caption{The density $\rho_\downarrow(x)$ of the minority species in the initial state~\eqref{Eq:InitialState} with the shifted Fermi sea of integers~\eqref{Eq:Is_bosons} in the Bose gas. The state is constructed with $N=100$ bosons at unit average filling for a number of interaction strengths.}
\label{Fig:Bosons_Shifted}
\end{figure}

\begin{figure}[t]
\begin{center}
\includegraphics[trim=0 0 0 25,clip,width=0.6\textwidth]{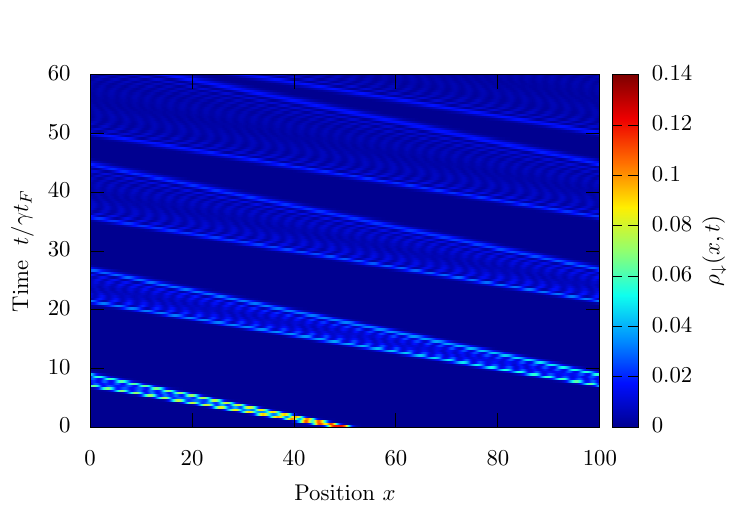}
\end{center}
\caption{Nonequilibrium dynamics of $\rho_\downarrow(x,t)$ in the Yang-Gaudin Bose gas starting from the initial state~\eqref{Eq:InitialState} with the shifted Fermi sea of integers~\eqref{Eq:Is_bosons}. The state is constructed for $N=100$ bosons at unit average density with interaction strength $c=10$.}
\label{Fig:NoneqIs_bosons}
\end{figure}

Finally, let us construct the state~\eqref{Eq:InitialState} in the Yang-Gaudin Bose gas with the shifted configuration of integers:
\begin{align}
\{ I \}^{(b)}_s = \{ -N+1, -N + 2, \ldots, 0 \}.
\label{Eq:Is_bosons}
\end{align}
As in the Fermi gas, we obtain the wave packet of spin excitations shown in Fig.~\ref{Fig:Bosons_Shifted}. As the initial state has a non-zero average momentum, under time-evolution the wave packet propagates to the left, whilst spreading, see Fig.~\ref{Fig:NoneqIs_bosons}. The spreading of the Bose gas is also suppressed compared to the Fermi sea of integers, in analogy with the explanation for the Fermi gas.

\section{Localized spin excitations in the weakly interacting Yang-Gaudin Fermi gas}
\label{App:Weakly}

\begin{figure}
\begin{center}
\includegraphics[trim=170 75 80 165,clip,width=0.45\textwidth]{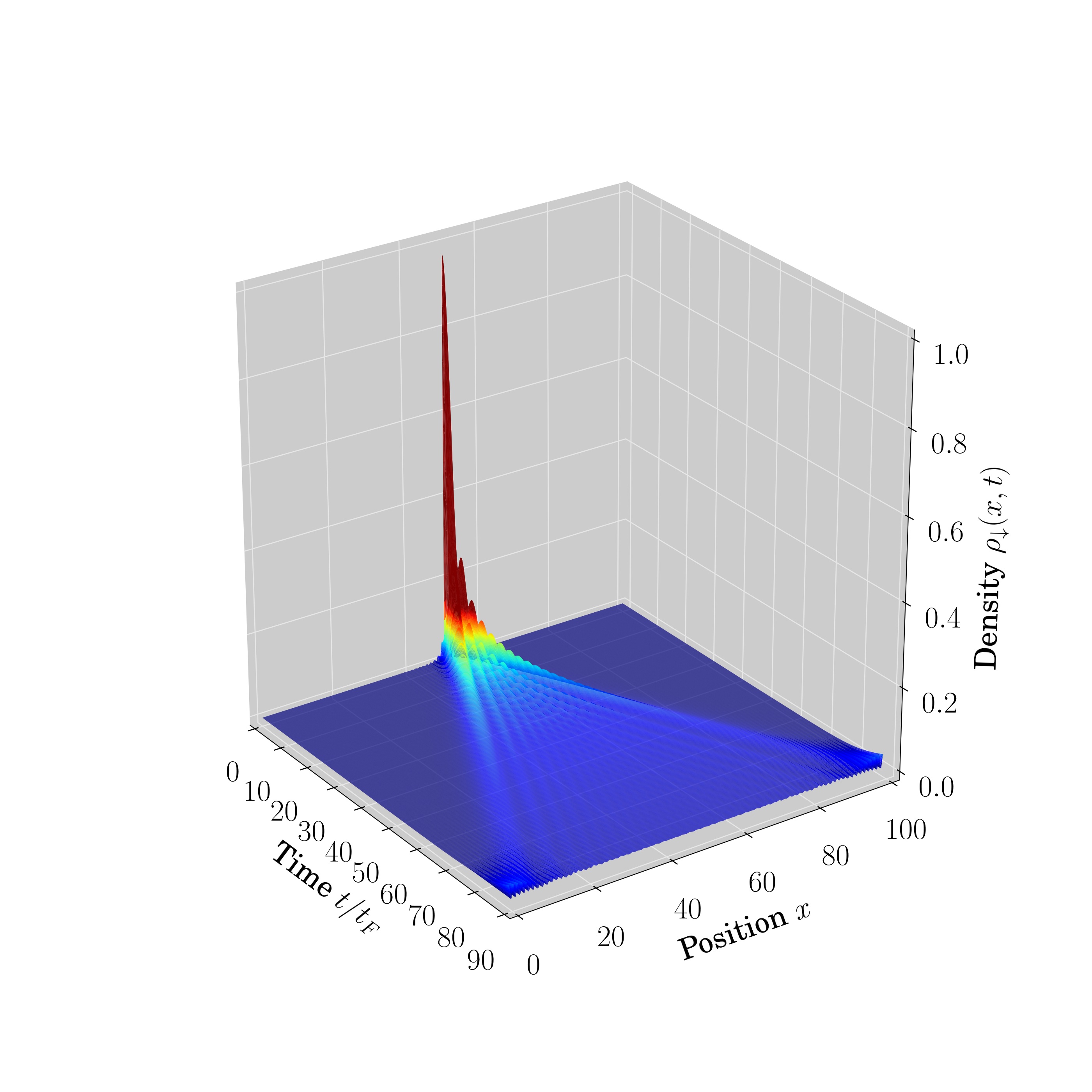} 
\includegraphics[trim=170 75 80 165,clip,width=0.45\textwidth]{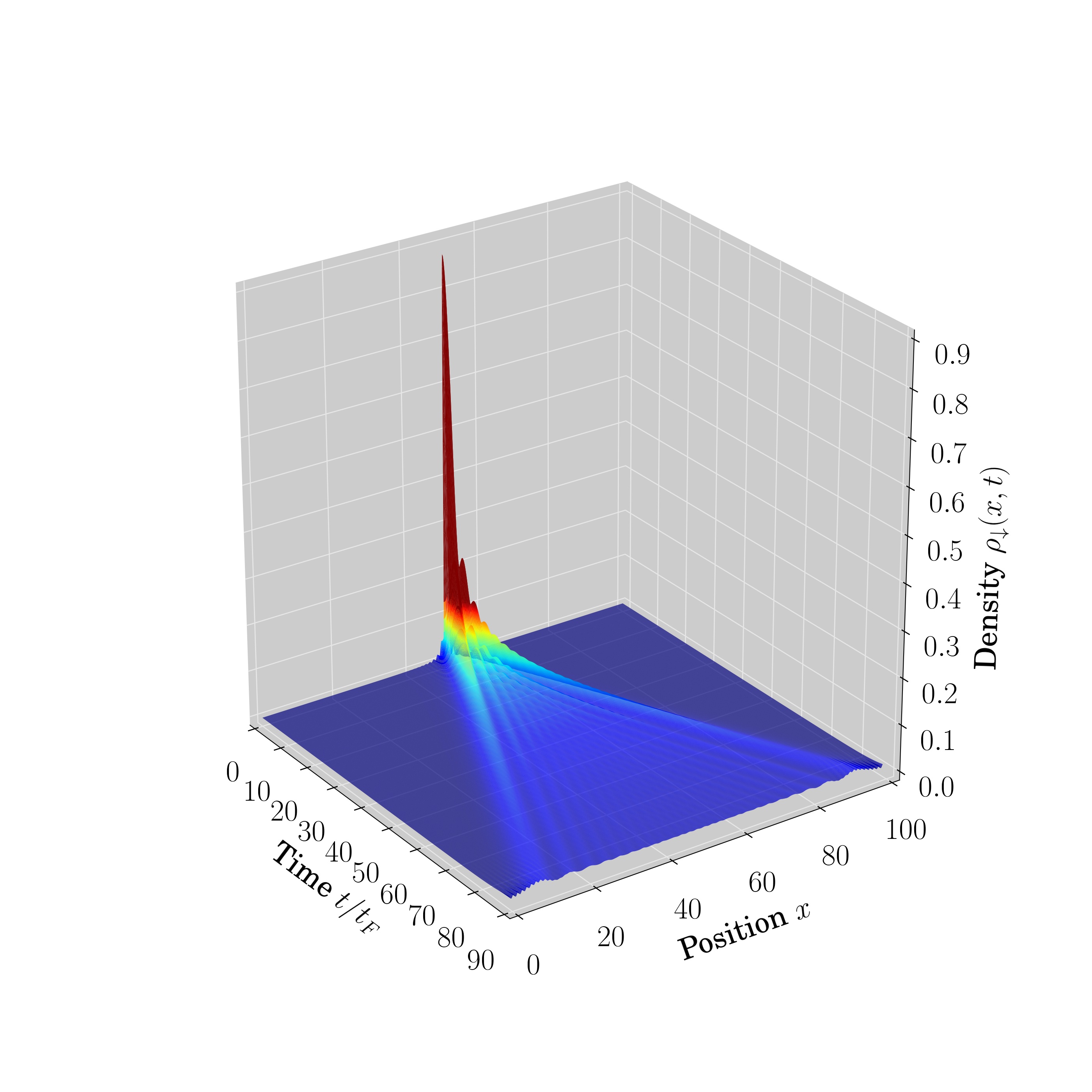}
\end{center}
\caption{Time-evolution of the minority species density $\rho_\downarrow(x,t)$ starting from the initial state~\eqref{Eq:InitialState} with the Fermi sea of integers~\eqref{Eq:If}. The state is constructed with $N=100$ fermions at unit average density and interaction strength (left) $c=0.1$ and (right) $c=0.5$. We see that the light cone is similar in the two cases, consistent with Eq.~\eqref{Eq:SpinWaveMassFermionsWeakCoupling}.}
\label{Fig:Fermions_WeakCoupling}
\end{figure}

In this appendix, we briefly consider the weak coupling limit of the Yang-Gaudin Fermi gas. By interpreting the initial state~\eqref{Eq:InitialState} with the Fermi sea configuration of integers~\eqref{Eq:If} as a linear superposition of spin excitations, we can gain some intuition for the behavior of the light cone. At weak coupling, the velocity of spin excitations with small momentum $P \ll \varrho$ is known~\cite{mcguire1965interacting} 
\begin{align}
v_s = 2 P \Bigg( 1 - \frac{\gamma^2}{\pi^4}  \Bigg) + O(\gamma^3).
\label{Eq:SpinWaveMassFermionsWeakCoupling}
\end{align}
As a result, we should expect the light cone at small values of the interaction parameter $\gamma$ to be approximately interaction independent (the interaction-dependent terms in Eq.~\eqref{Eq:SpinWaveMassFermionsWeakCoupling} are sub-leading). Indeed, we see this behavior in Fig.~\ref{Fig:Fermions_WeakCoupling}, where time-evolution of the minority species $\sigma = \, \downarrow$ for small interaction strengths $\gamma = 0.1,\, 0.5$ is presented for a system of $N=100$ particles at average unit density. In both these cases, by the time that excitations propagate around half the system (reaching the boundaries as plot), the density within the light cone is essentially constant and the initial localized wave packet of spin excitations has completely dissolved. 

\section{Relation to a simple spin flip state}
\label{App:spinflip}

Let us now address the following question: does the initial state~\eqref{Eq:InitialState} have a significant overlap with some simple-to-understand state, or are we constructing a state which is unlikely to be physically realized, even partially, in experiment? In order to address this question, we consider a simple state that contains a spatially localized spin flip excitation\footnote{We note that other choices of state could be made, which could lead to significant improvement or decline of the overlap with the initial state~\eqref{Eq:InitialState}.}
\begin{align}
|sf\rangle = \frac{1}{\cal N} \int {\rm d} x\, f(x) e^{iQx} \Psi^\dagger_\downarrow(x)\Psi^{\phantom\dagger}_\uparrow(x) | \Omega \rangle,
\label{Eq:spinflipstate}
\end{align}
where $f(x)$ defines the shape of the spin flip excitation, $|\Omega\rangle$ is an $N$-particle eigenstate of the one-component gas and $Q$ is a center-of-mass momentum for the spin flip excitation. Each of these constituents can be chosen to maximize the overlap of the spin flip state $|sf\rangle$ with the initial state~\eqref{Eq:InitialState}; the scalar product ${\cal C} = \langle \{ I \}_0|sf\rangle$ of the initial state with the spin flip excitation will give us a quantitative feel for how `close' the two states are. We can compute the scalar product ${\cal C}$ with knowledge of the matrix elements of the spin flip operator $\Psi^\dagger_\downarrow(x) \Psi_\uparrow(x)$, provided in Ref.~\cite{pozsgay2012on}. 

\subsection*{Appendix D.1. Yang-Gaudin Fermi Gas}

We fix $f(x)$ in~\eqref{Eq:spinflipstate} to reproduce the density, $\rho_\downarrow(x)$, of the minority species in the initial state~\eqref{Eq:InitialState} (as shown in Fig.~\ref{Fig:InitialDensityFermions}). We set $|\Omega\rangle$ to be the ground state of the one-component Fermi gas. Table~\ref{Tab:OverlapFermions} presents the overlap of the initial state~\eqref{Eq:InitialState} with the spin flip state~\eqref{Eq:spinflipstate} at a number of different interaction strengths. $Q \in (2\pi/R)\mathbb{Z}$ is chosen to maximize the overlap. For all interaction strengths, the spin flip state has significant overlap with the initial state of the main text, especially at larger values of $\gamma$. Importantly, states similar to $|sf\rangle$ can be prepared experimentally~\cite{palzer2009quantum}, so we should expect that our initial state~\eqref{Eq:InitialState} captures at least some features of the nonequilibrium dynamics observed in experiments.

\begin{table}[h]
\begin{center}
\begin{tabular}{ccc}
\hline
\hline
~~$\gamma$~~ &~~$QL/2\pi$~~& ~~Overlap $|{\cal C}|$~~ \\
\hline
2 & 3 & 0.3231\\
5 & 5 & 0.4260 \\
10 & 6 & 0.4788 \\
20 & 6 & 0.5052 \\
50 & 6 & 0.5175 \\
100 & 7 & 0.5208 \\
500 & 7 & 0.5233 \\ 
\hline
\hline
\end{tabular}
\caption{{\bf Fermions:} Overlap of the initial state~\eqref{Eq:InitialState} with the localized spin flip excitation~\eqref{Eq:spinflipstate} in the Yang-Gaudin Fermi gas with momentum $Q$, chosen to maximize the overlap. We choose $|\Omega\rangle$ to be the free fermion state with the Fermi sea of integers. The shape of the spin flip excitation, $f(x)$ in Eq.~\eqref{Eq:spinflipstate}, is chosen to reproduce the density of the impurity, see Fig.~\ref{Fig:InitialDensityFermions}. Here we have used $N=L=30$.}
\label{Tab:OverlapFermions}
\end{center}
\end{table} 

\subsection*{Appendix D.2. Yang-Gaudin Bose Gas}

As with the fermionic case, we consider the overlap of the initial state~\eqref{Eq:InitialState} with a simple spin flip excitation~\eqref{Eq:spinflipstate}. In Table~\ref{Tab:OverlapBosons} we present the overlap for a range of interaction strengths; once again, we see that the initial state~\eqref{Eq:InitialState} has a considerable overlap (albeit smaller than in the fermionic case) with simple spin flip states which may be prepared experimentally~\cite{palzer2009quantum}. 

\begin{table}[h]
\begin{center}
\begin{tabular}{ccc}
\hline
\hline
~~$\gamma$~~ & ~~$QL/2\pi$~~ &~~Overlap $|{\cal C}|$~~ \\
\hline
1 & 2 & 0.2002 \\ 
5 & 4 & 0.2926 \\ 
10 & 5 & 0.3353 \\ 
25 & 6 & 0.3573 \\ 
50 & 6 & 0.3621 \\ 
100 & 6 & 0.3640 \\ 
500 & 6 & 0.3653 \\ 
\hline
\hline
\end{tabular}
\caption{{\bf Bosons:} Overlap of the initial state~\eqref{Eq:InitialState} with the localized spin flip excitation~\eqref{Eq:spinflipstate} in the Yang-Gaudin Bose gas with momentum $Q$, chosen to maximize the overlap. We choose $|\Omega\rangle$ to be the one-component state with the Fermi sea of integers shifted one position to the left relative to the ground state. The shape of the spin flip excitation, $f(x)$ in Eq.~\eqref{Eq:spinflipstate}, is chosen to reproduce the density of the impurity, see Fig.~\ref{Fig:InitialDensityBosons}. Here we use $N = L = 30$. The computed overlap is only weakly dependent on the system size at fixed density.}
\label{Tab:OverlapBosons}
\end{center}
\end{table} 

\section{Example computations with $N = 500-1000$ particles}
\label{App:LargeN}

\begin{figure}
\begin{center}
\includegraphics[width=0.45\textwidth]{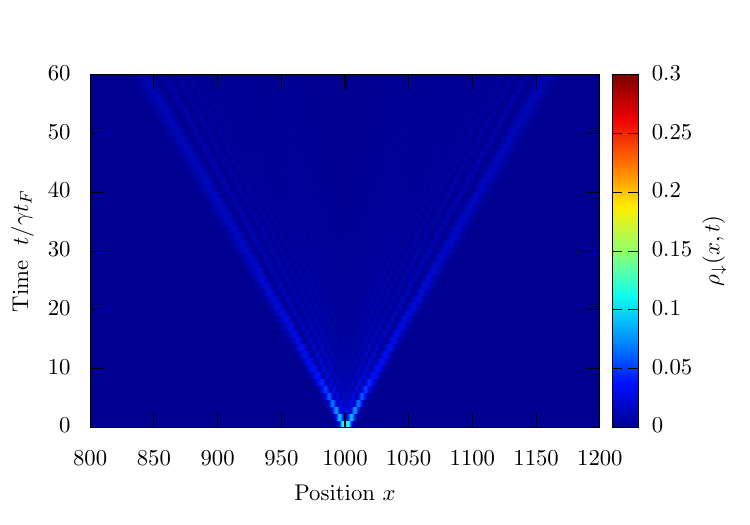}
\includegraphics[width=0.45\textwidth]{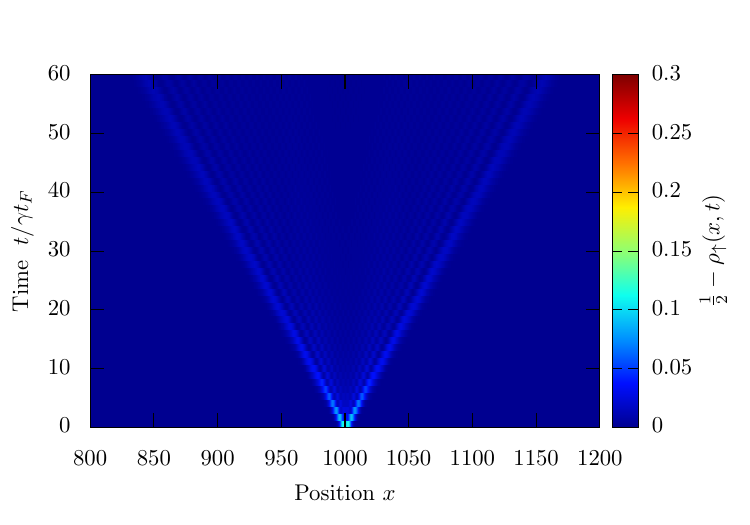}
\end{center}
\caption{{\bf Fermions:} Time-evolution of the density profile $\rho_\sigma(x,t)$ with (left panel) $\sigma =\ \downarrow$; (right panel) $\sigma =\ \uparrow$ for $N=1000$ fermions on the length $L=2000$ ring with interaction strength $c=50$. }
\label{Fig:Fermion_NLarge}
\end{figure}

\begin{figure}
\begin{center}
\includegraphics[width=0.6\textwidth]{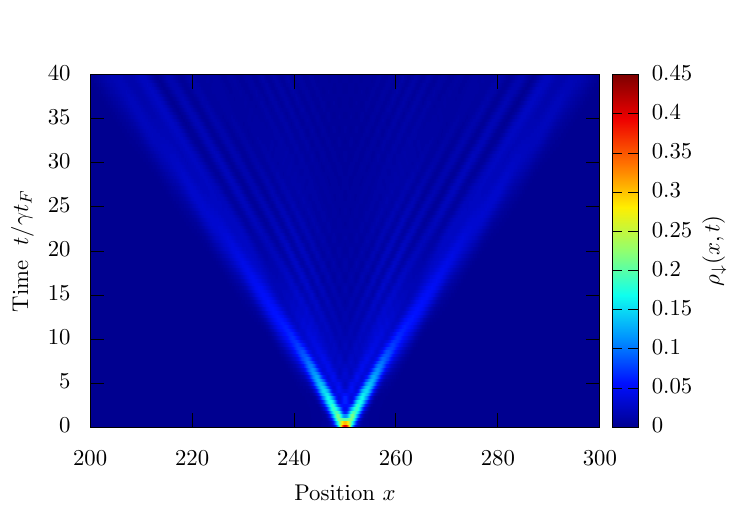} 
\end{center}
\caption{{\bf Bosons:} Time-evolution of the density profile $\rho_\downarrow(x,t)$ for $N=500$ bosons at unit filling $\varrho = 1$ with interaction strength $c=50$.}
\label{Fig:Boson_NLarge}
\end{figure}

In this appendix we provide some sample data for large numbers, $N=500-1000$, of particles. We first consider the time-evolution of the densities $\rho_\sigma(x,t)$ for initial state~\eqref{Eq:InitialState} of the Fermi gas with $N=1000$ particles at average density $\varrho = 1/2$ and $\gamma=100$. This is shown in Figure.~\ref{Fig:Fermion_NLarge}, where we see essentially the same behavior as smaller system sizes, cf. Figs.~\ref{Fig:FermionMajorityc50} and~\ref{Fig:Rescale}. In the second case, we consider the time-evolution of $\rho_\downarrow(x,t)$ for the $N=500$ particle initial state~\eqref{Eq:InitialState} of the Bose gas at unit average density $\varrho = 1$ and interaction strength $\gamma=50$. This shown in Fig.~\ref{Fig:Boson_NLarge}, which is similar to the data presented at smaller system sizes, cf. Fig.~\ref{Fig:Boson_cdep}(a), including the double light cone structure visible at long times. 

\vfill\eject

\section*{References}

\bibliographystyle{iopart-num}
\bibliography{bib_inits}

\end{document}